\documentclass[11pt]{article}

\newcommand{\bb}[0]{$\bullet$}

\usepackage{amstext}
\usepackage{enumerate}
\usepackage{citesort}
\usepackage{epsfig}

\def\blackslug{\hbox{\hskip 1pt \vrule width 4pt height 8pt depth 1.5pt
  \hskip 1pt}}
\def\myendproof{\quad\blackslug\lower 8.5pt\null}

\newenvironment{proof}{{\bf Proof. }}{\myendproof}
\newtheorem{theorem}{Theorem}[section]
\newtheorem{lemma}[theorem]{Lemma}
\newtheorem{corollary}[theorem]{Corollary}

\newtheorem{observation}[theorem]{Observation}
\newtheorem{example}{Example}
\newtheorem{problem}{Problem}

\newcommand{\tree}{\Psi} 
\newcommand{\closure}{\sigma}

\newcommand{\g}{G}
\newcommand{\hh}{H}
\newcommand{\ggg}{{\cal G}}
\newcommand{\gb}{{\cal H}}
\newcommand{\BB}{{\cal B}}
\newcommand{\CC}{{\cal C}}
\newcommand{\DD}{{\cal D}}
\newcommand{\EE}{\Phi}
\newcommand{\FF}{{\cal F}}
\newcommand{\LL}{{\cal L}}
\newcommand\MM{{\cal M}}
\newcommand{\NN}{{\cal N}}
\newcommand{\TT}{{\cal S}}
\newcommand{\PP}{{\cal P}}
\newcommand{\WW}{{\cal W}}
\newcommand{\UU}{{\cal U}}
\newcommand{\VV}{{\cal V}}
\newcommand{\rest}[2]{#1{\diamondsuit}#2}
\newcommand{\acf}{{\rm ACF}}
\newcommand{\acfv}{\Lambda_{\rm v}}
\newcommand{\acff}{\Lambda_{\rm f}}

\newcommand{\PT}{{\rm pert}}
\newcommand{\SK}{{\rm ske}}
\newcommand{\Fam}{{\rm fam}}
\newcommand{\Subset}{{\rm sub}}

\newcommand{\node}{{\rm node}}
\newcommand{\side}{{\rm side}}
\newcommand{\settle}{{\rm done}}
\newcommand{\depth}{{\rm dep}}
\newcommand{\ext}{{\rm ext}}
\newcommand{\low}{{\rm low}}
\newcommand{\Image}{{\rm img}}
\newcommand{\EF}{{\rm xfam}}
\newcommand{\ES}{{\rm xsub}}
\newcommand{\ESK}{{\rm xske}}
\newcommand{\post}{{\rm post}}
\newcommand{\rank}{{\rm rank}}

\newcommand{\inputsize}{I}

\title{Common-Face Embeddings of Planar Graphs\thanks{
The preliminary form of this paper appeared in the proceedings
of the 10th Annual ACM-SIAM Symposium on Discrete Algorithms,
1999, pp. 195-204.}}
\author{Zhi-Zhong Chen\thanks{Department of Mathematical Sciences,
		Tokyo Denki University, Hatoyama, Saitama 350-0394, Japan.
		Email: chen@r.dendai.ac.jp.} \and
	Xin He\thanks{Department of Computer Science and Engineering, State
                University of New York at Buffalo, Buffalo, NY 14260, USA.
		Email: xinhe@cse.buffalo.edu.
		Research supported in part by NSF Grant CCR-9912418.} \and
	Ming-Yang Kao\thanks{Department of Computer Science, Yale University,
		New Haven, CT 06520, USA. Email: kao-ming-yang@cs.yale.edu.
		Research supported in part by NSF Grant CCR-9531028.}}
\date{}

\begin{document}
\maketitle

\begin{abstract}
Given a planar graph $\ggg$ and a sequence ${\CC}_1,\ldots,{\CC}_q$,
where each ${\CC}_i$ is a family of vertex subsets of $\ggg$, we wish
to find a plane embedding of $\ggg$, if any exists, such that for each
$i\in\{1,\ldots,q\}$, there is a face $F_i$ in the embedding whose
boundary contains at least one vertex from each set in $\CC_i$.  This
problem has applications to the recovery of topological information
from geographical data and the design of constrained layouts in VLSI.
Let $\inputsize$ be the input size, i.e., the total number of vertices
and edges in $\ggg$ and the families ${\CC}_i$, counting multiplicity. 
We show that this problem is NP-complete in general. We also show that
it is solvable in $O(\inputsize\log \inputsize)$ time for the special case
where for each input family ${\CC}_i$, each set in ${\CC}_i$ induces a
connected subgraph of the input graph $\ggg$. Note that the classical
problem of simply finding a planar embedding is a further special case
of this case with $q=0$. Therefore, the processing of the additional
constraints ${\CC}_1,\ldots,{\CC}_q$ only incurs a logarithmic factor
of overhead.
\end{abstract}

\section{Introduction}\label{sec:intro} It is a fundamental problem in
mathematics
(e.g., see \cite{Frederickson95,kh97,kaofhr94,Mohar96,NR97,Tollis1996})
to embed a graph into a given surface while optimizing
certain objectives required by applications. (Throughout this paper,
a graph may have multiple edges and selfloops but a simple graph 
always has neither.)
A graph is {\it  planar} if it can be embedded on the plane so
that any pair of edges can only intersect at their endpoints; a {\it 
plane} graph is a planar one together with such an embedding.  A
classical variant of the problem is to test whether a given graph is
planar and in case it is, to find a planar embedding. This planarity
problem can be solved in linear time sequentially
{\cite{bl76,BM99,Mohar96}} and efficiently in parallel
{\cite{RamachandranR94}}.  

In this paper, we initiate the study of the following new planarity
problem.  Let $\ggg$ be a planar graph.
Let $\MM$ be a sequence ${\CC}_1,\ldots,{\CC}_q$, where each ${\CC}_i$ is
a family of vertex subsets of $\ggg$.  A plane embedding
${\EE}$ of $\ggg$ {\em satisfies} $\CC_i$ if the boundary of some face
in ${\EE}$ contains at least one vertex from each set in $\CC_i$.  
${\EE}$ {\em satisfies}
$\MM$ if it satisfies all $\CC_i$.  $\ggg$ {\it  satisfies} $\MM$ if
$\ggg$ has an embedding that satisfies $\MM$.

\begin{problem}[the common-face embedding (CFE) problem]\rm\  

\begin{itemize}
\item {\bf Input}: A planar graph $\ggg$ and a sequence $\MM$ of
	families of vertex subsets of $\ggg$.
\item
{\bf Question}: Does $\ggg$ satisfy $\MM$?
\end{itemize}
\end{problem}

Let $\inputsize$ be the input size, i.e., the total number of vertices 
and edges in $\ggg$ and the families $\CC_i$, counting multiplicity.
We first show that the CFE problem is NP-complete in general. Then,
for the special case where each vertex subset in each $\CC_i$ induces
a connected subgraph of $\ggg$, we give
an $O(\inputsize\log{\inputsize})$-time algorithm which can actually find
a plane embedding satisfying $\MM$, if any exists.
Note that the classical problem of simply finding a planar
embedding is a further special case of this special case with
$q=0$. Therefore, the processing of the additional constraints
${\CC}_1,\ldots,{\CC}_q$ only incurs a logarithmic factor of overhead.

The CFE problem arises naturally from topological inference
{\cite{CGP98}}.  For instance, in the conference version of this paper
\cite{chkao.map.scp}, a less general and less efficient variant of our
algorithm for the special case has been employed to design fast
algorithms for reconstructing maps from scrambled partial data in
geometric information systems {\cite{chkao.map.scp}}. In this
application {\cite{Ege91,ES93,EET76,GPP95,PS94,PSV97}}, each vertex
subset in $\MM$ describes a recognizable geographical feature and each
face in a planar embedding represents a geographical region.  Each
family in $\MM$ is a set of features that are known to be near each
other, i.e., surrounding the same region (on the boundary of the same face).  
Similarly, our algorithm
for the special case can compute a constrained layout of VLSI modules
\cite{GJ79}, where each vertex subset consists of the ports of a
module, and each subset family specifies a set of modules that are
required to be close to each other {\cite{chkao.map.scp}}.

To the best of our knowledge, the conference version of this paper is
the first to investigate the CFE problem {\cite{chkao.map.scp}}.  
A related problem has been studied in the context of speeding up the
computation of Steiner trees and
minimum-concave-cost network flows \cite{EMV1987,Provan1988,BM1988}.
Given a planar graph $G=(V,E)$ and a set of special vertices $S
\subseteq V$, the pair $(G,S)$ is called {\it  $k$-planar} if all the
vertices in $S$ are on the boundaries of at most $k$ faces of a planar
embedding of $G$.  Bienstock and Monma \cite{BM1988} showed that
testing $k$-planarity is NP-complete if $k$ is part of the input but
takes linear time for any fixed $k$.

The remainder of this paper is organized as follows.
Section \ref{sec:basic} proves the NP-completeness result and formally
states the main theorem on the CFE algorithm (Theorem~\ref{thm_main}).
Sections~\ref{sec:3-con} through \ref{sec:2-con} prove the main
theorem by detailing the algorithm for the key cases where $\ggg$ is
(1) triconnected, (2) disconnected, (3) connected, or (4) biconnected,
respectively. The triconnected case is the base case in that 
the other cases are eventually reduced to it. For this reason, 
this case is analyzed before the other cases. Section~\ref{sec_open}
concludes this paper with some directions for further research.

\section{Basics and the main results}\label{sec:basic}

\subsection{Basic definitions}
Let $\g$ be a graph. $|\g|$ denotes the {\em size} of $\g$, i.e.,
the total number of vertices and edges in $\g$. $\VV(\g)$ denotes
the vertex set of $\g$. If $\g$ is a plane graph,
then $\FF(\g)$ denotes the set of faces of $\g$.

A set $U$ is {\em $\g$-local} if $U \subseteq \VV(\g)$. A family $\CC$
of sets is {\em $\g$-local} if every set in $\CC$ is $\g$-local.

For a subset $U$ of $\VV(\g)$, the {\em subgraph of $\g$ induced by $U$}
is the graph $(U, E_U)$ where $E_U$ consists of all edges $e$ of $\g$
whose endpoints both belong to $U$; $G-U$ denotes the subgraph of $\g$
induced by $\VV(\g) - U$.

A {\em cut} vertex of $\g$ is one whose removal increases the number of
connected components in $\g$; a {\em block} of $\g$ is a maximal subgraph
of $\g$ with no cut vertex.  Let $\tree(\g)$ denote the forest whose vertices
are the cut vertices and the blocks of $\g$ and whose edges are those
$\{v, B\}$ such that $v$ is a cut vertex of $\g$, $B$ is a block of $\g$,
and $v \in \VV(B)$.  Note that $\tree(\g)$ is a tree if $\g$ is connected.

$\g$ is {\em biconnected} if it is connected and it has at least two
vertices but no cut vertex. $\g$ is {\em triconnected} if it is
biconnected, it has at least three vertices, and the removal of any
two vertices cannot disconnect it.

The {\em size} of a set $S$, denoted by $|S|$, is the number of elements
in $S$. The {\em size} of a family $\CC$ of sets, denoted by $|\CC|$,
is $\sum_{S}|S|$ where $S$ ranges over all sets in $\CC$.
The {\em size} of a sequence $\MM$ of families of sets, denoted by $|\MM|$,
is $\sum_{\CC}|\CC|$ where $\CC$ ranges over all families in $\MM$. 

\subsection{An NP-completeness result} 
\begin{theorem}\label{NPC}
The CFE problem is NP-complete.
\end{theorem}
\begin{proof}
We reduce the SATISFIABILITY problem \cite{GJ79} to the CFE problem.
Let $\phi$ be a CNF formula over variables $x_1,\ldots,x_n$ with $n \geq 2$. Let
$C_1,\ldots,C_m$ be the clauses of $\phi$, each regarded as the set of
literals in it.  We construct a simple biconnected planar graph
$\ggg=(V_1\cup V_2,E)$ as follows.  $V_1=\{x_1, \ldots, x_n\}
\cup\{\bar{x}_1, \ldots, \bar{x}_n\}\cup \{c_1, \ldots,
c_m\}$. $V_2=\{u_0, \ldots, u_n\}$.  For each $x_i$, $\ggg$ contains
edges $\{u_{i-1}, x_i\}$, $\{x_i, u_i\}$, $\{u_{i-1}, \bar{x}_i\}$,
$\{\bar{x}_i, u_i\}$. The only other edges of $\ggg$ are
$\{u_0,c_1\}$, $\{c_1,c_2\}$, $\{c_2,c_3\}$, \ldots,
$\{c_{m-1},c_m\}$, $\{c_m, u_n\}$, $\{u_n, u_0\}$.  Let $\MM$ be the
sequence $\{\{c_1\},C_1\},\ldots,\{\{c_m\},C_m\}$.  Observe that in
every plane embedding ${\EE}$ of $\ggg$, (1) the cycle
$c_1,\ldots,c_m,u_n,u_0$ forms the boundary of some face $F$ and (2)
for $i=1,\ldots,n$, exactly one of $x_i$ and $\bar{x}_i$ is on the
boundary of the face other than $F$ whose boundary contains the path
$c_1,\ldots,c_m$. Also, for every set $S \subseteq \{x_1, \ldots, x_n\}
\cup \{\bar{x}_1, \ldots, \bar{x}_n\}$ with $|S\cap \{x_i,
\bar{x}_i\}|=1$ for all $i=1,\ldots,n$, $\ggg$ has a plane embedding
where the boundary of some face contains the path $c_1,\ldots,c_m$ and
the vertices in $S$.  Therefore, $\phi$ is satisfiable if and only if
$\ggg$ satisfies $\MM$.
\end{proof}

\newcommand{\maintime}{$O(\inputsize\log\inputsize)$}

\subsection{The main theorem}
Although the input to the CFE problem is a planar graph $\ggg$, 
it is easy to see that $\ggg$ satisfies a given sequence $\MM$
if and only if its underlying simple graph (i.e., the simple graph obtained from $\ggg$
by deleting multiple edges and selfloops) satisfies the same $\MM$. 
Thus throughout the rest of this paper, unless explicitly stated otherwise,
$\ggg$ and $\MM$ always denote 
the input simple graph and the input sequence to our algorithm
for the CFE problem, respectively. Also, $\inputsize$ always denotes
$|\ggg| + |\MM|$, i.e., the size of the input to our algorithm.

The next theorem is the main theorem of this paper.  In light of this
theorem, the remainder of the paper assumes that every vertex subset
of $\ggg$ in $\MM$ induces a connected subgraph of $\ggg$.

\begin{theorem}\label{thm_main}
If every vertex subset in $\MM$ induces a connected subgraph
of $\ggg$, then the CFE problem can be solved in \maintime\ time.
\end{theorem}
\begin{proof} 
We consider three special cases:
\begin{itemize}
\item
{\it Case M1}: $\ggg$ is connected.
\item
{\it Case M2}: $\ggg$ is biconnected.
\item 
{\it Case M3}: $\ggg$ is triconnected.
\end{itemize}
In~\S\ref{sec:3-con}, Theorem~\ref{thm_m3} solves Case M3 of the CFE
problem faster than the desired time bound.  In \S\ref{sec:0-con},
Theorem~\ref{thm_m1_reduction} reduces this theorem to Case M1.  In
\S\ref{sec:1-con}, Theorem~\ref{thm_m2_reduction} reduces Case M1 to
Case M2.  In \S\ref{sec:2-con}, Theorem~\ref{thm_m2} uses
Theorem~\ref{thm_m3} to solve Case M2 of the CFE problem within the
desired running time.  This theorem follows from
Theorems~\ref{thm_m1_reduction}, \ref{thm_m2_reduction}, and
\ref{thm_m2}.
\end{proof}

As mentioned in Section~\ref{sec:intro}, Case M3 is the base case, 
meaning that the other cases are eventually reduced to it. So, 
the next section describes an algorithm for this case.

\section{Solving Case M3 where $\ggg$ is triconnected.}\label{sec:3-con}
This section assumes that $\ggg$ is triconnected.  Then, $\ggg$ has a
unique combinatorial embedding up to the choice of the exterior face
\cite{NC88,Whi33}. Thus, the CFE problem reduces in linear time to
that of finding all the faces in the embedding whose boundaries
intersect every set in some $\CC_i$. The naive algorithm takes $\Theta(|\ggg||\MM|)$
time. We solve the latter problem more efficiently by
recursively solving Problem~\ref{prob_acf} defined below.

Throughout this section, for technical convenience, the vertices of a
plane graph are indexed by distinct positive integers. The faces are
indexed by positive integers or $-1$.  The faces indexed by positive
integers have distinct indices and are called the {\it  positive}
faces.  Those indexed by $-1$ are the {\it  negative} faces.

Let $\gb$ be a plane graph.  A {\it  vf-set} of $\gb$ is a set of
vertices and positive faces in $\gb$.  A {\it  vf-family} of $\gb$ is a family
of vf-sets of $\gb$. A {\it  vf-sequence} of $\gb$ is a sequence of vf-families
of $\gb$. For a vf-family $\DD=\{S_1,\ldots,S_d\}$ of $\gb$, we define
$\acff(\gb,\DD)$ and $\acf(\gb,\DD)$ as follows:
\begin{enumerate}
\item
$\acfv(\gb,\DD) = \cap^d_{i=1} S_i\cap \VV(\gb)$.
\item
$\acff(\gb,\DD)$ is the set of positive faces $F$ of $\gb$ such that for
each $S_i\in\DD$, $F$ is a face in $S_i$ or its boundary intersects
$S_i-\acfv(\gb,\DD)$.
\item  $\acf(\gb,\DD)=\acfv(\gb,\DD)\cup\acff(\gb,\DD)$.
\end{enumerate}

\begin{problem}[the all-common-face (ACF) problem]\rm\ 
\label{prob_acf}

\begin{itemize}
\item
{\bf Input}: A plane graph $\gb$ and a vf-sequence $\NN$ of $\gb$.
\item
{\bf Output}: $\acf(\gb,\DD_1),\ldots,\acf(\gb,\DD_q)$ where
		$\DD_1,\ldots,\DD_q$ are the vf-families in $\NN$.
\end{itemize}
\end{problem}

Throughout the rest of this section, $\gb$ and $\NN$ always denote 
the input graph and the input sequence to our algorithm for the ACF problem,
respectively. 

To solve the ACF problem recursively, $\gb$ need not be simple or
triconnected. Furthermore, those faces that are indexed by $-1$ are
ruled out as final output during recursions.  To solve the problem
efficiently, each vertex in $\acfv(\gb,\DD_i)$ is meant as a succinct
representation of all the faces whose boundaries contain that vertex.
Similarly, the positive faces in the input $\DD_i$ and the output are
represented by their indices.

The next observation relates the CFE problem and the ACF problem.
\begin{observation}\label{obs_acf_cfe}
Let the faces of $\ggg$ be indexed by positive integers.  Then, the
output to the CFE problem is ``yes" if and only if for all $\CC_i$,
$\acf(\ggg,\CC_i)\neq\emptyset$.
\end{observation}

Section \ref{subsec_counting} proves a counting lemma useful for
analyzing the time complexity of our algorithms for the ACF problem.
Section~\ref{sec_reduce} provides a technique for simplifying $\gb$
during recursions.  Section~\ref{sec_acf_alg} uses this technique to
recursively solve the ACF problem without increasing the total size of
the subproblems.

\subsection{A counting lemma}\label{subsec_counting}

\begin{lemma}\label{lem_tri_count}\

\begin{enumerate}
\item\label{lem_tri_count_1}
Let $v_1$ and $v_2$ be distinct vertices in $\ggg$.  Let $F_1$ and $F_2$
be distinct faces in $\ggg$.  Then, both $v_1$ and $v_2$ are on the
boundaries of both $F_1$ and $F_2$ if and only if $v_1$ and $v_2$ form
a boundary edge of both $F_1$ and $F_2$.
\item\label{lem_tri_count_2}
Given a set $U$ of vertices in $\ggg$, there are $O(|U|)$ faces in $\ggg$
whose boundaries each contain at least two vertices in $U$.
\item\label{lem_tri_count_3}
Given a set $\PP$ of faces in $\ggg$, there are $O(|\PP|)$ vertices in
$\ggg$ which are each on the boundaries of at least two faces in
$\PP$.
\end{enumerate}
\end{lemma}
\begin{proof} We prove the statements separately as follows.

Statement~\ref{lem_tri_count_1}. This statement immediately follows
from the condition that $\ggg$ is triconnected with no multiple edges.

Statement~\ref{lem_tri_count_2}. Since $\ggg$ has no multiple edges, $\ggg$
contains $O(|U|)$ edges between distinct vertices in $U$.  Then, this
statement follows from Statement~\ref{lem_tri_count_1} and the fact
that an edge in a simple plane graph can be a boundary edge of at most two
faces.

Statement \ref{lem_tri_count_3}. If $\ggg$ has at most three vertices,
the statement holds trivially. Otherwise, the statement follows from
Statement~\ref{lem_tri_count_2} and the fact that the dual of $\ggg$ is
also a simple triconnected plane graph \cite{Ore67}.
\end{proof}

\begin{corollary}\label{cor_tri_count}
If $\gb$ is simple and triconnected, then the output of the ACF
problem has size $O(|\NN|)$.
\end{corollary}
\begin{proof}
This corollary follows from Lemma~\ref{lem_tri_count}(\ref{lem_tri_count_2}).
\end{proof}

\subsection{Simplifying $\gb$ over a vf-set}\label{sec_reduce}
To solve the ACF problem efficiently, we simplify the input graph $\gb$ by
removing unnecessary edges and vertices as follows.

\begin{figure}[h]
\centerline{\psfig{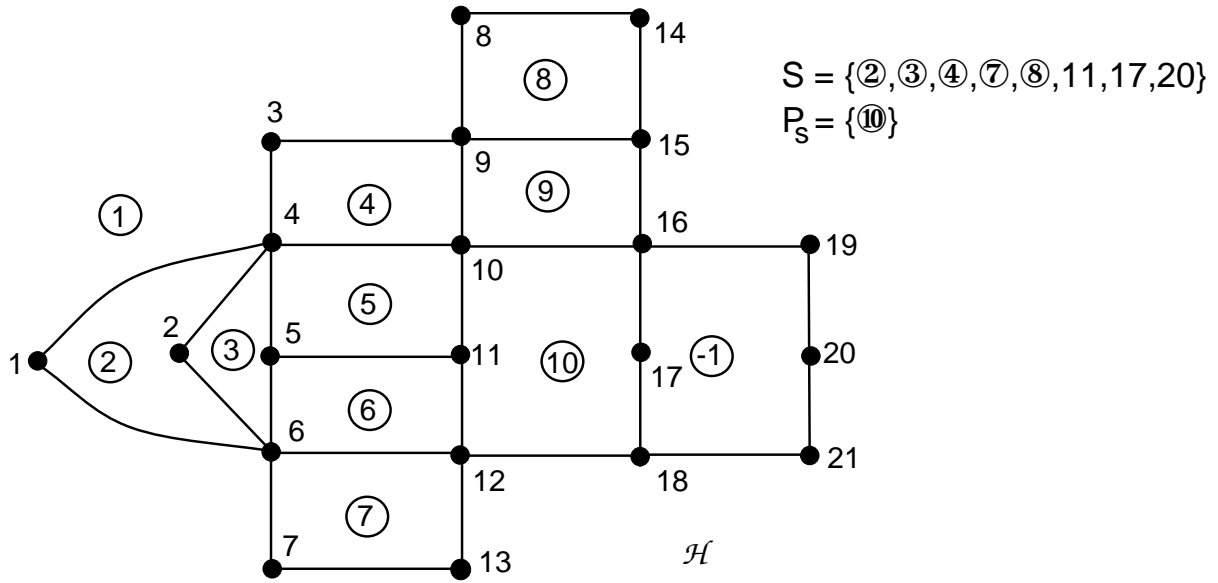}}
\caption{This is an example of a graph $\gb$, a vf-set $S$, and
$\PP_S$, where a number in a circle is the index of the corresponding
face.}
\label{kao-fig2}
\end{figure}

\begin{figure}
\centerline{\psfig{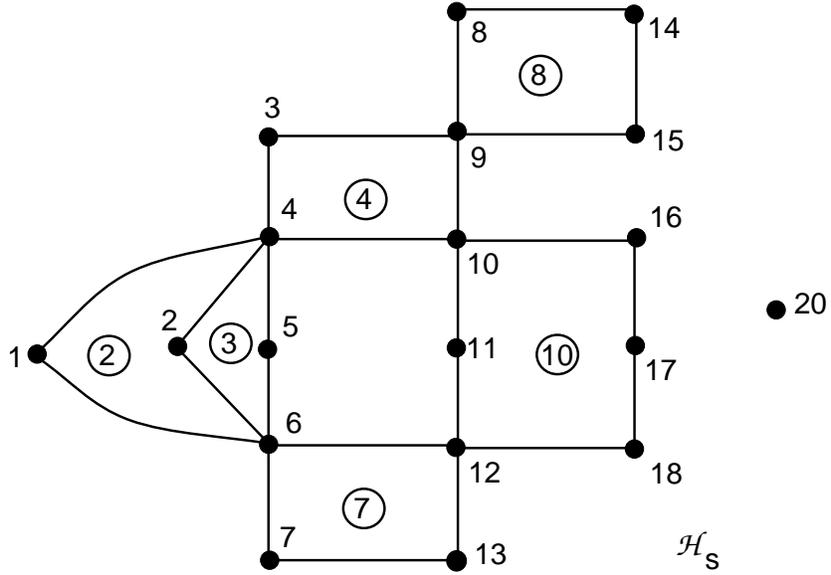}}
\caption{This is the graph $\gb_S$ for the example 
of $\gb$ and $S$ in Figure~\protect{\ref{kao-fig2}}.}
\label{kao-fig3}
\end{figure}

For a vf-set $S$ of $\gb$, the plane graph $\rest{\gb}{S}$ of
$\gb$ constructed as follows is said to {\it  simplify} $\gb$ over $S$.
An example is illustrated in Figures~\ref{kao-fig2},
\ref{kao-fig3}, and \ref{kao-fig1}.  

\begin{figure}
\centerline{\psfig{figure=kao-fig1.epsi,height=3in}}
\caption{This is the graph $\rest{\gb}{S}$ for the example of $\gb$ and $S$ 
in Figure~\protect{\ref{kao-fig2}}.}
\label{kao-fig1}
\end{figure}

Let $\PP_S$ be the set of the positive faces in $\gb$ whose boundaries each
contain at least two distinct vertices in $S\cap \VV(\gb)$.
Let $\gb_S$ be the plane subgraph of $\gb$ (1) whose vertices are
those in $S\cap \VV(\gb)$ and the boundary vertices of the faces
in $(S\cap \FF(\gb))\cup\PP_S$ and (2) whose edges are the boundary edges
of the faces in $(S\cap \FF(\gb))\cup\PP_S$.
Note that $\gb_S$ inherits a plane embedding from $\gb$.

Let $U_3$ be the set of vertices which are of degree at least three in
$\gb_S$; note that each vertex in $U_3$ appears on the boundaries of
at least two faces in $(S\cap \FF(\gb)) \cup \PP_S$.
A {\it  compressible} path $P$ in $\gb_S$ is a maximal path, which may be a cycle, 
such that (1)
every internal vertex of $P$ appears only once in it, and (2) no
internal vertex of $P$ is in $S \cup U_3$. 
Note that by the choice of $U_3$, every internal vertex of a compressible path
is of degree 2 in $\gb_S$.
We use this property to further simplify $\gb_S$. Let $\rest{\gb}{S}$ be
the plane graph obtained from $\gb_S$ by replacing each compressible path
with an edge between its endpoints.  This edge is embedded by the same
curve in the plane as the path is.  For technical consistency, if a
compressible path forms a cycle and its endpoint is not in $S \cup U_3$,
then we replace it with a self-loop for the vertex of the cycle with
the smallest index.

Each vertex in $\rest{\gb}{S}$ is given the same index as in $\gb$.
Note that the closure of the interior of each face of $\rest{\gb}{S}$
is the union of those of several faces or just one in $\gb$.  Let $F$
be a face in $\rest{\gb}{S}$ and $F'$ be one in $\gb$.  Let $\closure$
(respectively, $\closure'$) denote the closure of the interior of $F$
(respectively, $F'$).  If $\closure=\closure'$, then $F$ and $F'$ are
regarded as the same face, and $F$ is assigned the same index in
$\rest{\gb}{S}$ as $F'$ is in $\gb$.  For technical conciseness, these
two faces are identified with each other.  If $\closure$ is the union
of the closures of the interiors of two or more faces in $\gb$, $F$ is
not the same as any face in $\gb$ and is indexed by $-1$. This
completes the definition of $\rest{\gb}{S}$.

\begin{lemma}\label{lem_rest_count}\

\begin{enumerate}
\item\label{lem_rest_count_1}
Given $\gb$ and $S$, we can compute $\rest{\gb}{S}$ in $O(|\gb|+|S|)$ time.
\item\label{lem_rest_count_3} Let $S'$ be a vf-set of $\rest{\gb}{S}$.  If $S'
\subseteq S$, then $\rest{\gb}{S'}=\rest{(\rest{\gb}{S})}{S'}$.
\item\label{lem_rest_count_2} If $\gb$ simplifies $\ggg$ over a vf-set
$S^*$ with $S \subseteq S^*$, then $|\rest{\gb}{S}|=O(|S|)$.
\end{enumerate}
\end{lemma}
\begin{proof}
Statements \ref{lem_rest_count_1} and \ref{lem_rest_count_3} are
straightforward. To prove Statement~\ref{lem_rest_count_2}, it
suffices to prove $|\rest{\ggg}{S}|=O(|S|)$ since by
Statement~\ref{lem_rest_count_3}, $\rest{\gb}{S}=\rest{\ggg}{S}$.

To bound the number of vertices in $\rest{\ggg}{S}$, let $\PP_S$
and $U_3$ be as specified in the definition of $\rest{\ggg}{S}$.  Let
$U_1$ be the set of vertices $v$ in $\rest{\ggg}{S}$ such that $v$ appears
on the boundary of exactly one face in $(S\cap \FF(\ggg))\cup\PP_S$.  Then,
$(S\cap\VV(\ggg)) \cup U_3 \cup U_1$ consists of all the vertices in
$\rest{\ggg}{S}$.  Note that $|U_1|\leq|(S\cap\FF(\ggg))\cup\PP_S|$. Also,
by Lemma~\ref{lem_tri_count}(\ref{lem_tri_count_3}),
$|U_3|=O(|(S\cap\FF(\ggg))\cup\PP_S|)$.  Consequently, since by
Lemma~\ref{lem_tri_count}(\ref{lem_tri_count_2}) $|\PP_S|=O(|(S\cap\VV(\ggg))|)$,
$|(S\cap\VV(\ggg)) \cup U_3 \cup U_1|=O(|S|)$ as desired.

To bound the number of edges in $\rest{\ggg}{S}$, we first examine the
multiple edges. Let $u$ and $v$ be adjacent vertices in
$\rest{\ggg}{S}$.  Let $X_{u,v}$ be the set of faces in
$(S\cap\FF(\ggg))\cup\PP_S$ whose boundaries contain both $u$ and $v$. Then,
$|X_{u,v}| \geq 1$.  By Lemma
\ref{lem_tri_count}(\ref{lem_tri_count_1}), $|X_{u,v}| \leq 2$.  If
$X_{u,v}=\{F\}$, then the two boundary paths of $F$ between $u$ and
$v$ may degenerate into at most two multiple edges between $u$ and $v$
in $\rest{\ggg}{S}$.  If $X_{u,v}=\{F_1,F_2\}$, then by the triconnectivity 
of $\ggg$, $F_1$ and $F_2$ share exactly one common boundary edge $e$,
which is also an edge in $\rest{\ggg}{S}$. 
Let $C_i$ be the boundary of $F_i$ without $e$.
$C_1$ and $C_2$ may degenerate into at most two multiple edges between
$u$ and $v$ in $\rest{\ggg}{S}$.  In summary, there are at most three
multiple edges between two vertices in $\rest{\ggg}{S}$.  Similarly, only
the boundary of a face in $S\cap\FF(\ggg)$ can degenerate into a self-loop
in $\rest{\ggg}{S}$; so, $\rest{\ggg}{S}$ has only $O(|S|)$ self-loops.
By Euler's formula, $\rest{\ggg}{S}$ has $O(|S|)$ edges as desired.
\end{proof}

\subsection{Algorithms for the ACF problem}
\label{sec_acf_alg}
Throughout this subsection, let $\DD_1,\ldots,\DD_q$ be the vf-families in $\NN$.
To solve the ACF problem recursively, we use simplification to reduce
the number of $\DD_i$ and the number of sets in each $\DD_i$.

For brevity, we define several notations.
For a vf-family $\DD$ of $\gb$, let $\rest{\gb}{\DD}=\rest{\gb}{(\cup_{S\in\DD}S)}$.
For a vf-sequence $\NN'$: $\DD'_1,\ldots,\DD'_p$ of $\gb$, 
let $\rest{\gb}{\NN'}=\rest{\gb}{(\DD'_1\cup\cdots\cup \DD'_p)}$.
For a vf-set $S^*$ of $\gb$ and a vf-family $\DD$ of $\gb$,
we say $\DD \leq S^*$ if $S \subseteq S^*$ for all $S\in\DD$.
For a vf-set $S^*$ of $\gb$, we say $\NN \leq S^*$ if $\DD_i \leq S^*$
for all $\DD_i$, $1\leq i\leq q$.

Lemmas~\ref{lem_reduce_bi_vf} and \ref{lem_reduce_many_vf} below
reduce to $1$ the number of $\DD_i$ in $\NN$ in the ACF problem.

\begin{lemma}\label{lem_reduce_bi_vf}
Assume $q \geq 2$.  Let $\NN_\ell=\DD_1,\ldots,\DD_{\lceil
q/2\rceil}$ and $\NN_r=\DD_{\lceil q/2\rceil+1},\ldots,\DD_q$.  Let
$\gb_\ell=\rest{\gb}{\NN_\ell}$ and $\gb_r=\rest{\gb}{\NN_r}$.
\begin{enumerate}
\item\label{lem_reduce_bi_vf_3} Given $\gb$ and $\NN$, we can compute
$\gb_\ell$ and $\gb_r$ in $O(|\gb|+|\NN|)$ total time.
\item\label{lem_reduce_bi_vf_1}
For $1 \leq i \leq \lceil q/2 \rceil$,
$\rest{\gb}{\DD_i}=\rest{\gb_\ell}{\DD_i}$.  Similarly, for $\lceil q/2
\rceil+1 \leq i \leq q$, $\rest{\gb}{\DD_i}=\rest{\gb_r}{\DD_i}$.
\item\label{lem_reduce_bi_vf_2} If $\gb$ simplifies $\ggg$ over a
vf-set $S^*$ with $\NN \leq S^*$, then $|\gb_\ell| = O(|\NN_\ell|)$ and
$|\gb_r|=O(|\NN_r|)$.
\end{enumerate}
\end{lemma}
\begin{proof}
The three statements follow from those of Lemma \ref{lem_rest_count},
respectively.
\end{proof}

\begin{lemma}\label{lem_reduce_many_vf}
Assume $q \geq 1$.  Let $\gb_i=\rest{\gb}{\DD_i}$.
\begin{enumerate}
\item\label{lem_reduce_many_vf_1}
$\acf(\gb,\DD_i)=\acf(\gb_i,\DD_i)$.
\item\label{lem_reduce_many_vf_2} If $\gb$ simplifies $\ggg$ over a
vf-set $S^*$ with $\NN \leq S^*$, then $|\gb_i|=O(|\DD_i|)$.
\item\label{lem_reduce_many_vf_3} If $\gb$ simplifies $\ggg$ over a
vf-set $S^*$ with $\NN \leq S^*$, then given $\gb$ and $\NN$, we can
compute all $\gb_i$ in $O(|\gb|+|\NN|\log(q+1))$ total time.
\end{enumerate}
\end{lemma}
\begin{proof} We prove the statements separately as follows.

Statement~\ref{lem_reduce_many_vf_1}. The proof is straightforward.
Note that a positive face in $\gb_i$ is also a positive face in $\gb$
and that a negative face in $\gb_i$ combines one or more faces not in
$\acf(\gb,\DD_i)$. 

Statement~\ref{lem_reduce_many_vf_2}. The proof follows from
Lemma~\ref{lem_rest_count}(\ref{lem_rest_count_2}).

Statement~\ref{lem_reduce_many_vf_3}.  The graphs $\gb_i$ can be
computed by applying Lemma~\ref{lem_reduce_bi_vf} recursively with
$O(\log (q+1))$ iterations.  By
Lemma~\ref{lem_reduce_bi_vf}(\ref{lem_reduce_bi_vf_3}), the first
iteration takes $O(|\gb|+|\NN|)$ time.  By
Lemmas~\ref{lem_reduce_bi_vf}(\ref{lem_reduce_bi_vf_2}) and
\ref{lem_reduce_bi_vf}(\ref{lem_reduce_bi_vf_3}), each subsequent
iteration takes $O(|\NN|)$ time.  By
Lemma~\ref{lem_rest_count}(\ref{lem_rest_count_3}), the constant
coefficient in the $O(|\NN|)$ term does not accumulate over
recursions.
\end{proof}

Lemma~\ref{lem_D_reduce_d} below solves the ACF problem with only one
$\DD_i$ in $\NN$. 
\begin{lemma}\label{lem_D_reduce_d}
Let $\DD=\{S_1,\ldots,S_d\}$ be a vf-family of $\gb$ where $d\geq 1$.
Let $\DD'_\ell=\{S_1,\ldots,S_{\lceil d/2\rceil}\}$
and $\DD'_r=\{S_{\lceil d/2\rceil+1},\ldots,S_d\}$.  Let
$\gb_\ell=\rest{\gb}{\DD'_\ell}$; $\gb_r=\rest{\gb}{\DD'_r}$; and 
$\DD''=\{\acf(\gb_\ell,\DD'_\ell),\acf(\gb_r,\DD'_r)\}$.
\begin{enumerate}
\item\label{lem_D_reduce_d_1}
$\acf(\gb,\DD) = \acf(\gb,\DD'')$.
\item\label{lem_D_reduce_d_2} If $\gb$ simplifies $\ggg$ over a vf-set
$S^*$ with $\DD \leq S^*$, then given $\gb$ and $\DD$, $\acf(\gb,\DD)$
can be computed in $O(|\gb|+|\DD|\log (d+1))$ time.
\end{enumerate}
\end{lemma}
\begin{proof} The statements are proved separately as follows.

Statement~\ref{lem_D_reduce_d_1}.  Note that
$\acf(\gb,\DD)=\acf(\gb,\{\acf(\gb,\DD'_\ell),\acf(\gb,\DD'_r)\})$ by a
straightforward case analysis. Then, as Lemma
\ref{lem_reduce_many_vf}(\ref{lem_reduce_many_vf_1}),
$\acf(\gb,\DD'_\ell)=\acf(\gb_\ell,\DD'_\ell)$ and
$\acf(\gb,\DD'_r)=\acf(\gb_r,\DD'_r)$.

Statement~\ref{lem_D_reduce_d_2}.  We compute $\acf(\gb,\DD)$
recursively via Statement~\ref{lem_D_reduce_d_1}. 
If $d=1$, then $\acf(\gb,\DD)=S_1$. 
If $d=2$, then $\acf(\gb,\DD)$ can be computed in $O(|\gb|)$ time in a
straightforward manner.
For $d > 2$, there are three stages:
\begin{enumerate}
\item
Compute $\gb_\ell$ and $\gb_r$ in $O(|\gb|+|D|)$ time in a straightforward
manner.
\item
Recursively compute $\acf(\gb_\ell,\DD'_\ell)$ and $\acf(\gb_r,\DD'_r)$.

\item Compute $\acf(\gb,\DD'')$ in $O(|\gb|)$ time in a straightforward
manner, which is $\acf(\gb,\DD)$ by Statement~\ref{lem_D_reduce_d_1}.
\end{enumerate}
This recursive computation has $\log d + O(1)$ iterations.  The
recursion at the top level takes $O(|\gb|+|\DD|)$ time.  Every
subsequent level takes $O(|\DD|)$ time since by
Lemma~\ref{lem_rest_count}(\ref{lem_rest_count_2})
$O(|\gb_\ell|)=O(|\DD'_\ell|)$ and $O(|\gb_r|)=O(|\DD'_r|)$.
Note that by
Lemma~\ref{lem_rest_count}(\ref{lem_rest_count_3}), the constant
coefficient in the $O(|\DD|)$ term does not accumulate over
recursions.
\end{proof}

The next theorem is the main result of this section.
\begin{theorem}\label{thm_m3}\

\begin{enumerate}
\item\label{thm_m3_1} Let $d$ be the maximum number of vf-sets in any
$\DD_i$ in $\NN$. If $\gb$ simplifies $\ggg$ over a vf-set $S^*$ with
$\NN \leq S^*$, then the ACF problem can be solved in $O(|\gb|+|\NN|
\log (d+q))$ time.
\item\label{thm_m3_2} Let $d$ be the maximum number of vertex sets in
any $\CC_i$ in $\MM$.  Case M3 of the CFE problem can be solved in
$O(|\ggg|+|\MM|\log (d+q))$ time.
\end{enumerate}
\end{theorem}
\begin{proof} 
Statement \ref{thm_m3_1} follows from Lemmas \ref{lem_reduce_many_vf}
and \ref{lem_D_reduce_d}.  Statement \ref{thm_m3_2} follows from
Observation~\ref{obs_acf_cfe}, Statement \ref{thm_m3_1}, and the fact that
$\ggg$ has a unique combinatorial embedding computable in linear time
\cite{NC88,Whi33}.
\end{proof}

In \S\ref{subsec:ana}, the algorithm for Case M2 of the CFE problem
calls Theorem \ref{thm_m3}(\ref{thm_m3_2}) to solve subproblems in which
some $S \in \CC_i$ may consist of a single edge $\{u,v\}$. For such
subproblems, we replace $S$ by $\{u\}$ and $\{v\}$ and then apply
Theorem~\ref{thm_m3}(\ref{thm_m3_2}).

\section{Reducing  Theorem~\ref{thm_main} to Case M1 where $\ggg$ is connected.}
\label{sec:0-con} 
Let $\ggg_1,\ldots,\ggg_k$ be the connected components of $\ggg$.
Let $\CC_1,\ldots,\CC_q$ be the families in $\MM$.
A family $\CC_h$ in $\MM$ is {\em global} if for every $i\in\{1,\ldots,k\}$,
$\CC_h$ is not $\ggg_i$-local.  Let $\hh$ be an edge-labeled graph defined 
as follows.  The vertices of $\hh$ are $1,\ldots,k$.  For each
global $\CC_h$, $\hh$ contains a cycle $C$ possibly of length 2 where
(1) the vertices of $C$ are those $i\in\{1,\ldots,k\}$ such that
some set in $\CC_h$ is $\ggg_i$-local and (2) the edges of $C$ are
all labeled $h$. See Figures \ref{fig:chen-fig1}(1) through
\ref{fig:chen-fig1}(3) for an example of $\ggg$, $\MM$ and $H$.

\begin{figure}[t]
\centerline{\psfig{figure=chen-fig1.epsi,height=4.5in}}
\caption{(1) This is a simple disconnected graph $\ggg$ with six connected components $\ggg_1$
through $\ggg_6$ where $\VV(\ggg_1)=\{1, \ldots, 8\}$,
$\VV(\ggg_2)=\{9, \ldots, 13\}$, $\VV(\ggg_3)=\{21,22\}$,
$\VV(\ggg_4)=\{16, \ldots, 20\}$, $\VV(\ggg_5)=\{23, \ldots, 25\}$ and
$\VV(\ggg_6)=\{14,15\}$.
(2) This is a sequence $\MM$ of families of vertex subsets of $\ggg$ where
$\CC_6$ and $\CC_7$ are  $\ggg_1$-local but the rest families are global.
(3) This is the graph $H$ constructed from $\ggg$ and $\MM$.
(4) These are the sequences constructed for $\ggg_1$ through $\ggg_6$, respectively.}
\label{fig:chen-fig1}
\end{figure}

\begin{observation}\label{obs_connectH}
Let $\hh_1,\ldots,\hh_\ell$ be the connected components of $\hh$.  For
each $\hh_j$, let $\ggg'_j$ be the subgraph of $\ggg$ formed by all
$\ggg_i$ with $i\in \VV(\hh_j)$. Let $\MM'_j$ be
the sequence of all $\ggg'_j$-local families in $\MM$.  Then,
$\ggg$ satisfies $\MM$ if and only if every $\ggg'_j$ satisfies $\MM'_j$.
\end{observation}

By Observation~\ref{obs_connectH}, we may assume that $\hh$ is connected.  Let
$B_1, \ldots, B_p$ be the blocks of $\hh$.  Then, for each
global $\CC_h$, exactly one $B_j$ contains all the edges labeled $h$.
For every $B_j$, let ${\UU}_j=\cup_h \CC_h$ where $h$ rangers over
all labels on the edges of $B_j$.
For each $\ggg_i$, let $\MM_i$ be the sequence
consisting of the $\ggg_i$-local families in $\MM$ as well as the
families ${\UU}_{j, i}=\{U\in {\UU}_j~|~U$ is $\ggg_i$-local$\}$ for
all $B_j$ with $i \in \VV(B_j)$. See Figure \ref{fig:chen-fig1}(4)
for an example of $\MM_1,\ldots,\MM_6$ constructed from $\ggg$, $\MM$ and $H$ in
Figures \ref{fig:chen-fig1}(1) through \ref{fig:chen-fig1}(3).

\begin{lemma}\label{0-con:reduce}
$\ggg$ satisfies $\MM$ if and only if every $\ggg_i$ satisfies $\MM_i$.
\end{lemma}

\begin{proof} The two directions are proved as follows.

$(\Longrightarrow)$ Let ${\EE}$ be an embedding of $\ggg$ satisfying
$\MM$.  Let ${\EE}_i$ be the restriction of ${\EE}$ to $\ggg_i$. For
each $\ggg_i$, our goal is to prove that ${\EE}_i$ satisfies $\MM_i$.
First, ${\EE}_i$ satisfies each $\ggg_i$-local family in $\MM$.  Let
$B_j$ be a block of $\hh$ with $i\in B_j$.  We next
prove that ${\EE}_i$ satisfies ${\UU}_{j,i}$.  Let
$i,i_1,\ldots,i_\ell$ be the vertices of $B_j$.
We claim that $\ggg$ has no cycle $C$ such that at least
one but not all of $\ggg_i$, $\ggg_{i_1},\ldots,\ggg_{i_\ell}$ are inside
$C$ in ${\EE}$.  To prove by contradiction, assume that such $C$
exists.  Then, some $\ggg_x$ with $1\leq x\leq k$ contains $C$.
However, by the construction of $\hh$, no connected component of
$\hh-\{x\}$ contains all of $i$, $i_1,\ldots,i_\ell$,
contradicting the fact that $B_j$ is a
block of $\hh$. Thus, the claim holds. Therefore, the boundary of some
face $F$ in ${\EE}$ intersects each of
$\ggg_i,\ggg_{i_1},\ldots,\ggg_{i_\ell}$. Since $F$ must be unique, the boundary 
of $F$ intersects every set in $\CC_h$ for every $\CC_h$ in $\MM$ such that 
the sets in $\CC_h$ fall into two or more of $\ggg_i$, $\ggg_{i_1}$, $\ldots$, 
$\ggg_{i_\ell}$. Hence, the boundary of $F$ intersects every set
in ${\UU}_j$.  Consequently, ${\EE}_i$ satisfies ${\UU}_{j,i}$.

$(\Longleftarrow)$ Let ${\EE}_i$ be an embedding of $\ggg_i$
satisfying $\MM_i$. We construct an embedding of $\ggg$ satisfying
$\MM$ as follows.  First, consider a block $B_j$ of $\hh$.  Let
$i_1,\ldots,i_\ell$ be the vertices of $B_j$.
Let $\ggg'_j$ be the subgraph of $\ggg$ formed by
$\ggg_{i_1}$, \ldots, $\ggg_{i_\ell}$. Let $\MM'_j$ be the sequence
consisting of ${\UU}_j$ and the $\ggg_{i_x}$-local families in $\MM$
for $x=1,\ldots,\ell$.  We can assume that the boundary of the exterior
face of ${\EE}_{i_x}$ intersects every set in ${\UU}_{j,i_x}$.  By
identifying the exterior faces of ${\EE}_{i_1}$, \ldots,
${\EE}_{i_\ell}$, we can combine the embeddings into an embedding
${\EE}'_j$ of $\ggg'_j$ satisfying $\MM'_j$.  Next, we utilize
$T=\tree(\hh)$ to combine ${\EE}'_1$, \ldots, ${\EE}'_p$ into a
single embedding of $\ggg$.  First, root $T$ at a block of $\hh$.  For
a leaf $B_{j_1}$ in $T$, let $\ggg_i$ and $B_{j_2}$ be the parent and
grandparent of $B_{j_1}$ in $T$, respectively.  Let ${\LL}_{i,1}$
(respectively, ${\LL}_{i,2}$) be the restriction of ${\EE}'_{j_1}$
(respectively, ${\EE}'_{j_2}$) to $\ggg_i$. Note that ${\EE}_i$,
${\LL}_{i,1}$, and ${\LL}_{i,2}$ are topologically equivalent up to 
the choice of their exterior face. Thus, ${\EE}'_{j_1}$
(respectively, ${\EE}'_{j_2}$) can be obtained as follows: For every
vertex $i' \not= i$ of $B_{j_1}$ (respectively,
$B_{j_2}$), put a suitable embedding ${\LL}_{i'}$ of $\ggg_{i'}$ that
is topologically equivalent to ${\EE}_{i'}$ \ into a suitable face
$F_{i'}$ of ${\EE}_i$.  This gives an embedding of those $\ggg_x \in
\{\ggg_1,\ldots,\ggg_k\}$ with $x\in \VV(B_{j_1})\cup\VV(B_{j_2})$. We replace
${\EE}'_{j_2}$ with this embedding, replace $B_{j_2}$ with the union
of $B_{j_1}$ and $B_{j_2}$, and delete $B_{j_1}$ from $T$. Afterwards,
if $\ggg_i$ becomes a leaf of $T$, then we further delete it from
$T$. We repeat this process until $T$ is a single vertex, at which
time we obtain an embedding of $\ggg$ satisfying $\MM$.
\end{proof}

\begin{theorem}\label{thm_m1_reduction}
Theorem~\ref{thm_main} holds if it holds for Case M1.
\end{theorem}
\begin{proof} 
The proof follows from Lemma~\ref{0-con:reduce} and the fact that
$\hh$ and the sequences $\MM_i$ above can be constructed from $\ggg$
and $\MM$ in $O(\inputsize)$ time.
\end{proof}

\section{Reducing Case M1 to Case M2 where $\ggg$ 
is biconnected.} \label{sec:1-con} 
This section assumes Case M1 where $\ggg$ is connected.
We also assume that $\ggg$ has at least two vertices;
otherwise, the problem is trivial.

Section~\ref{subsec:single} shows how to eliminate one cut vertex from $\ggg$;
iterating this elimination until $\ggg$ has no cut vertex gives us a reduction
from Case M1 to Case M2. However, this reduction is not efficient. 
Section~\ref{subsec:all} describes a more efficient reduction based on
a direct elimination of all cut vertices from $\ggg$.
Throughout the rest of this section, 
let $\CC_1,\ldots,\CC_q$ be the families in $\MM$.

\subsection{Eliminating one cut vertex}\label{subsec:single} Let $w$ be
a cut vertex of $\ggg$. Let $W_1$, \ldots, $W_k$ be the vertex sets of
the connected components of $\ggg-\{w\}$.  Let $\ggg_i$ be the
subgraph of $\ggg$ induced by $\{w\} \cup W_i$.
$\ggg_1,\ldots,\ggg_k$ are called the {\em augmented components} induced by
$w$.  For each $\CC_h$ in $\MM$, let $U_{h,1}$, \ldots, $U_{h,t_h}$ be
the sets in $\CC_h$ containing $w$; possibly $t_h=0$.  $\CC_h$ is 
{\em $w$-global} if for all $i\in\{1,\ldots,k\}$,
$\CC_h-\{U_{h,1},\ldots,U_{h,t_h}\}$ is not $\ggg_i$-local; otherwise,
$\CC_h$ is {\em $w$-local}.

\begin{observation}\label{obs_Shrink}\
 
\begin{enumerate}
\item\label{lem_shrink_1} Assume that $\CC_h-\{U_{h,1}, \ldots,
U_{h,t_h}\}$ is $\ggg_i$-local for some $\ggg_i$. Then, $\ggg$
satisfies $\MM$ if and only if $\ggg$ satisfies $\MM$ with $\CC_h$ replaced
by $(\CC_h-\{U_{h,1},\ldots,U_{h,t_h}\})$ ${\cup}$
$\{U_{h,1}{\cap}\VV(\ggg_i)$,\ldots,$U_{h,t_h}{\cap}\VV(\ggg_i)\}$.
\item\label{lem_shrink_2} Assume that $\CC_h$ is $w$-global.  Then,
$\ggg$ satisfies $\MM$ if and only if $\ggg$ satisfies $\MM$ with
$\CC_h$ replaced by $\CC_h - \{U_{h,1},\ldots,U_{h,t_h}\}$.
\end{enumerate}
\end{observation}
%\begin{proof} Let $\EE$ be a plane embedding of $\ggg$. Suppose that
%the boundary of a face $F$ in $\EE$ contains two vertices $u$ and $v$
%such that $u\not=w$, $v\not=w$, and $u$ and $v$ belong to two distinct
%augmented components $\ggg_i$ and $\ggg_j$, respectively. Then, starting at $u$,
%we can walk to $v$ along the boundary of $F$. Since $w$ is the only vertex
%common to both $\ggg_i$ and $\ggg_j$, we must pass through $w$. Thus, the boundary
%of $F$ contains the vertex $w$. This implies the observation.
%\end{proof}

By Observation~\ref{obs_Shrink}, we may assume that (1) each set in a $w$-global family
in $\MM$ does not contain $w$ and (2) each set in a family in $\MM$ is
$\ggg_i$-local for some $\ggg_i$.  Let $\hh$ be an edge-labeled graph
constructed as follows.  The vertices of $\hh$ are $1, \ldots, k$.
For each $w$-global family $\CC_h$, $\hh$ has a cycle $C$
possibly of length 2 where (1) the vertices of $C$ are those $i\in\{1,\ldots,k\}$
such that at least one set in $\CC_h$ is $\ggg_i$-local and (2) the
edges of $C$ are all labeled $h$. See Figures \ref{fig:chen-fig2}(1) through 
\ref{fig:chen-fig2}(3) for 
an example of $\ggg$, $\MM$ and $H$.

\begin{figure}[t]
\centerline{\psfig{figure=chen-fig2.epsi,height=4.2in}}
\caption{(1) This is a simple connected graph $\ggg$ with a cut vertex 2. It induces four augmented
components $\ggg_1$ through $\ggg_4$ with $\VV(\ggg_1)=\{1,2\}$,
$\VV(\ggg_2)=\{2, \ldots, 6\}$, $\VV(\ggg_3)=\{2, 7, \ldots,10\}$ and
$\VV(\ggg_4)=\{2, 11, \ldots, 21\}$.
(2) This is a sequence $\MM$ of families of vertex subsets of $\ggg$ 
    where only $\CC_2$ through $\CC_5$ are 2-global.
(3) This is the graph $H$ constructed from $\ggg$ and $\MM$.
(4) These are the sequences constructed for $\ggg_1$ through $\ggg_4$, respectively.}
\label{fig:chen-fig2}
\end{figure}

Note that Observation~\ref{obs_connectH}
still holds for this $\hh$ and the augmented components
$\ggg_1,\ldots,\ggg_k$. Thus, we may assume that $\hh$ is connected.
Let $B_1$, $\ldots$, $B_p$ be the blocks of $\hh$.  Clearly, for each
$w$-global family $\CC_h\in\MM$, exactly one block of $\hh$ contains
all the edges labeled $h$.  For each $B_j$, let ${\UU}_j=\cup_h\CC_h\cup\{\{w\}\}$
where $h$ ranges over all labels on the edges of $B_j$.
For each $\ggg_i$, let $\MM_i$ be the sequence 
consisting of the $\ggg_i$-local families in $\MM$ as well as the
families ${\UU}_{j,i}=\{U\in {\UU}_j~|~U$ is $\ggg_i$-local\}
for all $B_j$ with $i \in \VV(B_j)$. See Figure \ref{fig:chen-fig2}(4) for an
example of $\MM_1,\ldots,\MM_4$ constructed from $\ggg$, $\MM$ and $H$ in
Figures \ref{fig:chen-fig2}(1) through \ref{fig:chen-fig2}(3).

\begin{lemma}\label{1Reduce}
$\ggg$ satisfies $\MM$ if and only if every $\ggg_i$ satisfies
$\MM_i$.
\end{lemma}

\begin{proof} The two directions are proved as follows.

$(\Longrightarrow)$ The proof is the same as that of
Lemma~\ref{0-con:reduce} except that the claim therein now implies that
the boundary of some face $F$ in ${\EE}$ intersects each of
$\ggg_i-\{w\},\ggg_{i_1}-\{w\},\ldots,\ggg_{i_\ell}-\{w\}$.

$(\Longleftarrow)$ The proof is the same as that of
Lemma~\ref{0-con:reduce} except that ${\EE}'_{j_1}$ (respectively,
${\EE}'_{j_2}$) now can be obtained as follows:  For each vertex $i'\not= i$
of $B_{j_1}$ (respectively, $B_{j_2}$), put a
suitable embedding ${\LL}_{i'}$ of $\ggg_{i'}$ that is topologically
equivalent to $\EE_{i'}$ into a suitable face $F_{i'}$ of ${\EE}_i$,
and then identify the two occurrences of $w$.
\end{proof}

\subsection{Eliminating all cut vertices}\label{subsec:all}
Let $T=\tree(\ggg)$. A {\em block vertex} of $T$ is a vertex of $T$
that is a block of $\ggg$. Root $T$ at a block vertex and perform 
a post-order traversal of $T$. For each vertex $\gamma$ of $T$, let
$\post(\gamma)$ be the post-order number of $\gamma$
in the post-order traversal of $T$.

Let $W=\{w_1, \ldots, w_\ell\}$ be the set of cut vertices of $\ggg$ where
$\post(w_1)<\cdots<\post(w_\ell)$.  For each $v\in \VV(\ggg) - W$, let
$\post(v)=\post(B)$, where $B$ is the unique block of $\ggg$ with $v\in
\VV(B)$.  We may assume $\VV(\ggg)=\{1, \ldots, n\}$.  For each $v\in \VV(\ggg)$,
the {\em rank} of $v$, denoted by $\rank(v)$, is $(\post(v),v)$.  The rank of a
vertex $u$ is {\em lower} than that of another vertex $v$ if (1)
$\post(u)<\post(v)$ or (2) $\post(u)=\post(v)$ and $u<v$.  For each
$w_i\in W$, let $B_{i,1}$, \ldots, $B_{i,k_i}$ be the children of
$w_i$ in $T$. Let $B_{i,0}$ be the parent of vertex $w_i$ in $T$.

\begin{theorem}\label{thm_m2_reduction}
Theorem~\ref{thm_main} holds for Case M1 if it holds for Case M2.
\end{theorem}
\begin{proof}
It suffices to construct a sequence $\MM[B]$ for each block $B$ of
$\ggg$, with a total size of $O(\inputsize)$ in \maintime\ total time
over all the blocks of $\ggg$, such
that $\ggg$ satisfies $\MM$ if and only if every $B$ satisfies
$\MM[B]$.  To construct $\MM[B]$ based on Observation~\ref{obs_Shrink} 
and Lemma~\ref{1Reduce}, we process $w_1$, \ldots, $w_\ell$ one at a
time. During the processing of $w_i$, we construct $\MM[B_{i,j}]$
for all $j=1,\ldots,k_i$. Then, we delete $w_i$, $B_{i,1}$, \ldots,
$B_{i,k_i}$ from $T$. After processing $w_\ell$, we are left with the
root $B_{\ell,0}$ for which we then construct $\MM[B_{\ell,0}]$.

We use the following data structures. See Figure \ref{fig:chen-fig3}
for an example of some of the data structures
before processing the first cut vertex of $\ggg$.

\begin{figure}[t]
\centerline{\psfig{figure=chen-fig3.epsi,height=2.5in}}
\caption{(1) This is $\tree(\ggg)$ where $\ggg$ is the simple graph in Figure \ref{fig:chen-fig2}(1).
Here, $\VV(B_1)=\{1,2\}$, $\VV(B_2)=\{2,\ldots,6\}$, $\VV(B_3)=\{2,7,\ldots,10\}$, 
$\VV(B_4)=\{2,11,\ldots,14\}$, $\VV(B_5)=\{11,15,16\}$, $\VV(B_6)=\{14,17,\ldots,20\}$, 
$\VV(B_7)=\{20,21\}$.
The number to the left of each vertex $\gamma$ of $\tree(\ggg)$ is $\post(\gamma)$,
and the list to the right is $L(\gamma)$ before processing the first cut vertex of $\ggg$.
For visibility, each set $U$ in a pair in $L(\gamma)$ with
$U\cap W \neq \emptyset$ is divided into two parts via a semiclolon;
the first part consists of vertices in $U \cap W$ in the increasing order of their post-order numbers.
(2) These are the representatives in the union-find data structure before processing the
first cut vertex of $\ggg$.
(3) This is the array $A_1$ before processing the first cut vertex of $\ggg$.}
\label{fig:chen-fig3}
\end{figure}

\begin{enumerate}
\item During the construction, some
families in $\MM$ may be united, and we use a union-find data
structure to maintain a collection of disjoint dynamic subsets of
$\Delta=\{1,\ldots,q\}$.
(Recall that $q$ is the number of families in $\MM$.)
Each subset of $\Delta$ in the data structure
is identified by a {\em representative} member of the subset. For each
$h\in \Delta$, let $R(h)$ be the representative of the subset
containing $h$.  Initially, each $h\in\Delta$ forms a singleton
subset, and thus, $R(h)=h$. 
\item Each set $U$ in a family in $\MM$ is implemented as
a pair $({\WW}[U], {\TT}[U])$, where ${\WW}[U]$ is a
linked list, and ${\TT}[U]$ is a splay tree \cite{ST85}.  Initially,
${\WW}[U]$ consists of the vertices in $U\cap W$ in the increasing
order of their post-order numbers. ${\TT}[U]$ is initialized by inserting
the ranks of the vertices in $U-W$ into an empty splay tree.  A splay
tree supports the following operations in amortized logarithmic time
per operation: (1) insert a rank and (2) delete the ranks in a given range. 
\item A linked list $L[B]$, for each block $B$ of $\ggg$. Initially,
  each $L[B]$ consists of all pairs $(h,U)$ such that $h\in\Delta$,
  $U\in \CC_h$, $U$ is $B$-local, and $U\cap W=\emptyset$.
\item A linked list $L[w_i]$, for each $w_i\in W$. Initially,
  each $L[w_i]$ consists of all pairs $(h,U)$ such that $h\in\Delta$,
  $U\in \CC_h$, $w_i \in U$, and $i=\min\{j~|~w_j\in U\cap W\}$. 
\item An array $A_1[1..q]$ of integers. Initially,
  for each $h\in\Delta$, $A_1[h] = \max_{\gamma} \post(\gamma)$ where
  $\gamma$ ranges over all vertices of $T$ such that $L[\gamma]$ contains
  a pair $(h,*)$ with $*$ = ``don't care".
\item An array $A_2[1..q]$ of integers.
  Initially, for each $h\in\Delta$, $A_2[h]=0$.
\item An array $J[1..q]$ of linked lists of integers.
  Initially, for each $h\in\Delta$, $J[h]$ is empty.
\item A temporary array $Y[1..q]$ of integers.
\end{enumerate}

We maintain the following invariants immediately before processing each
$w_i$.  In particular, we initialize the above data structures so that
the invariants hold before $w_1$ is processed.  It takes $O(\inputsize)$
total time to initialize the data structures except the splay trees.
\begin{enumerate}
\item
  For each vertex $\gamma$ of $T$ and each pair
  $(h,U)\in L[\gamma]$, (1) ${\WW}[U]$ consists of the vertices
  in $U\cap\{w_i,\ldots,w_\ell\}$ in the increasing order of their
  post-order numbers,
  (2) the rank of each vertex of $U-\{w_i,\ldots,w_\ell\}$ is stored
	in ${\TT}[U]$, and
  (3) for every $w_j\in U\cap\{w_1,\ldots, w_{i-1}\}$,
	$\post(w_j)$ and $\rank(w_j)$ have been updated as
	$\post(B_{j,0})$ and $(\post(B_{j,0}), w_j)$, respectively.

\item
    For each block vertex $B$ of $T$ and each $(h,U)\in L[B]$, it holds that $h \in
    \Delta$, $U$ is $B$-local, and $U\cap \{w_i,\ldots,w_\ell\}=\emptyset$.

\item For each $j\in\{i,\ldots, \ell\}$ and each $(h,U)\in L[w_j]$, it holds that $h
  \in\Delta$, $w_j\in U$, and $j=\min\{x~|~i\leq x\leq \ell$ and $w_x\in U\}$.

\item\label{iff} For each $h\in\Delta$ with $R(h)=h$, let $\CC'_h=\{U~|$
there is a vertex $\gamma$ of $T$
such that $L[\gamma]$ contains a pair $(h',U)$ with $R(h')=h\}$.
Let $\MM'$ be the sequence of all families $\CC'_h$ such that
$h\in\Delta$ and $R(h)=h$. Let $\ggg'$ be the subgraph of $\ggg$
induced by $\cup_B \VV(B)$, where $B$ ranges over all the block vertices of
$T$. Then, $\ggg$ satisfies $\MM$ if and only if (1) $\ggg'$ 
satisfies $\MM'$ and (2) for each block $B$ of $\ggg$ that has been deleted
from $T$, $B$ satisfies $\MM[B]$.

\item For each $h\in\Delta$ with $R(h)=h$, $A_1[h] = \max_{\gamma}
  \post(\gamma)$ where $\gamma$ ranges over all vertices of $T$ such
  that $L[\gamma]$ contains a pair $(h',*)$ with $R(h')=h$.

\item For each $h\in\Delta$, $A_2[h]=0$ and $J[h]$ is empty.
\end{enumerate}

We process $w_i$ in the following stages W1 through W4. See Figure \ref{fig:chen-fig4}
for an example of some of the data structures after processing the first cut
vertex of $\ggg$.

\begin{figure}[t]
\centerline{\psfig{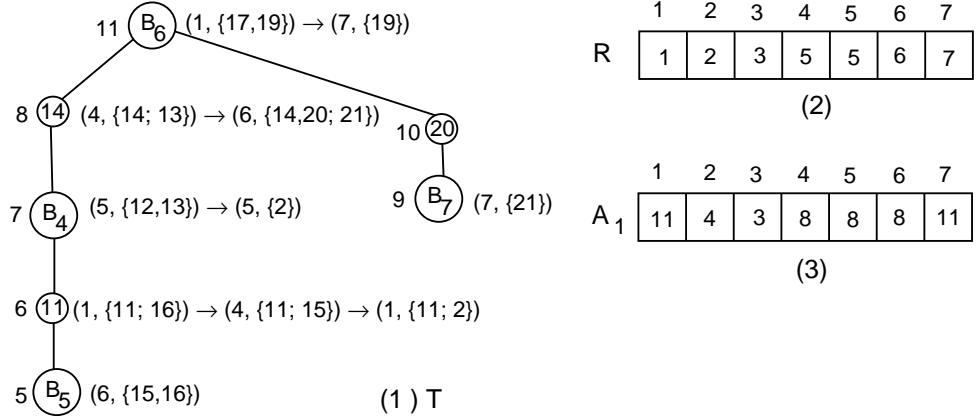}}
\caption{This is the data structure after processing the first cut vertex (i.e., the
vertex 2) of the graph in Figure \ref{fig:chen-fig4}(1).}
\label{fig:chen-fig4}
\end{figure}

\bb{} Stage W1 checks whether each related family is $w_i$-global as follows.
\begin{enumerate}
\item Compute $X = \{h\in\Delta~|~R(h)=h$, and for some $j\in\{1,\ldots,k_i\}$,
      $L[B_{i,j}]$ contains a pair $(h',*)$ with $R(h')=h\}$. ({\it  
      Remark.} For each $h\in\Delta-X$ with $R(h)=h$, the family 
      $\CC'_h-\{U~|~w_i\in U\}$ is $Q_i$-local, where $Q_i$ is the augmented
      component of $\ggg'$ induced by $w_i$ that is not among $B_{i,1}$,
      \ldots, $B_{i,k_i}$. See the fourth invariant for $\CC'_h$ and $\ggg'$.)

\item For each $h\in X$, set $Y[h]$ to be the number of integers $j \in
	\{1,\ldots,k_i\}$ such that $L[B_{i,j}]$ contains a pair
	$(h',*)$ with $R(h')=h$. ({\it  Remark.} For $h\in X$, $Y[h]\geq 1$.)

\item For each $h\in X$, perform the following:
\begin{enumerate} 
\item 
If $Y[h]=1$ and $A_1[h] \leq \post(w_i)$, then set $A_2[h]=j$ where $j$ is
 the unique integer in $\{1,\ldots,k_i\}$ such that $L[B_{i,j}]$ contains a pair
 $(h',*)$ with $R(h')=h$.  ({\it Remark.} 
 $\CC'_h - \{U~|~w_i\in U\}$ is $B_{i,j}$-local.)
\item 
Otherwise set $A_2[h]=-1$.  ({\it  Remark.}  $\CC'_h -
\{U~|~w_i\in U\}$ is $w_i$-global.)  \end{enumerate}
\end{enumerate}

\bb{} Stage W2 modifies each $U$ with $w_i \in U$ in each $w_i$-local family
based on Observation~\ref{obs_Shrink}(\ref{lem_shrink_1}) as follows.
\begin{enumerate}
\item For each $(h,U)\in L[w_i]$ with $A_2[R(h)]\geq 1$, let $j=A_2[R(h)]$,
	delete all vertices outside $\VV(B_{i,j})$ from $U$,
	and then insert $(h,U)$ to $L[B_{i,j}]$. Here, deleting
	all vertices outside $\VV(B_{i,j})$ from $U$ is done as follows:
	Delete $w_i$ from ${\WW}[U]$, delete all the ranks in the range 
	$[-\infty..(\post(B_{i,j}), 0)]$ and all the ranks in the range
	$[(\post(B_{i,j}), n+1)..\infty]$ from ${\TT}[U]$, and
	insert $(\post(B_{i,j}), w_i)$ to ${\TT}[U]$.

\item For each $(h,U)\in L[w_i]$ with $A_2[R(h)]=0$,
	perform the following: \\ ({\it Remark.} 
 	$\CC'_h - \{U~|~w_i\in U\}$ is $Q_i$-local.
	See the remark in Step 1 of Stage W1 for $Q_i$.)

\begin{enumerate}
\item 
Delete all vertices $v$ with $\post(v) < \post(w_i)$ from $U$ as follows: 
Delete $w_i$ from ${\WW}[U]$, delete all the ranks in the range 
$[-\infty..\rank(w_i)]$ from ${\TT}[U]$, and insert $(\post(B_{i,0}), w_i)$
to ${\TT}[U]$.  

\item 
If ${\WW}[U]=\emptyset$, i.e., $U$ has no cut vertex, then insert
$(h,U)$ to $L[B_{i,0}]$ and set
$A_1[R(h)]=\max\{\post(B_{i,0}),A_1[R(h)]\}$.

\item 
If ${\WW}[U]\not=\emptyset$, then find the first vertex $w_j$ in
${\WW}[U]$, insert $(h,U)$ to $L[w_j]$, and set
$A_1[R(h)]=\max\{\post(w_j), A_1[R(h)]\}$.  ({\it  Remark}. $j>i$.)
\end{enumerate}
\end{enumerate}

\bb{} Stage W3 modifies each $w_i$-global family based on
Observation~\ref{obs_Shrink}(\ref{lem_shrink_2}) as follows.
\begin{enumerate}
\item For each $h\in X$ with $A_2[h]=-1$, set $J[h] = \{j \in \{1, \ldots,
	k_i\}~|~L[B_{i,j}]$ contains a pair $(h',*)$ with $R(h')=h\}$.

\item For each $h\in X$ with $A_2[h]=-1$ and $A_1[h] > \post(w_i)$, insert
      0 to $J[h]$.

\item Set $\post(w_i)=\post(B_{i,0})$ and $\rank(w_i)=(\post(B_{i,0}), w_i)$.

\item Construct an edge-labeled graph $\hh_i$ as follows.  The vertices
	of $\hh_i$ are $0$, $1$, \ldots, $k_i$.  For each $h\in X$ with
	$A_2[h]=-1$, $\hh_i$ has a cycle possibly of length 2 whose
	vertices are the integers in $J[h]$ and whose edges are all
	labeled $h$.

\item For each block ${\BB}$ of $\hh_i$, find the labels $h_1,\ldots,h_t$ on 
the edges in ${\BB}$ and unite those subsets in the union-find data structure 
that have $h_1,\ldots,h_t$ as their representative, respectively; afterwards, 
for the representative $h_r$ of the resulting subset, further perform 
the following:
\begin{enumerate}
  \item\label{add_ai:1} Insert $(h_r, \{w_i\})$ to all lists $L[B_{i,j}]$ 
	such that $j\in \VV({\BB})$.

  \item\label{add_ai:2} If $0\in \VV({\BB})$, then set $A_1[h_r] = 
	\max\{\post(B_{i,0})$, $A_1[h_1]$, \ldots, $A_1[h_t]\}$.
	\end{enumerate}
\end{enumerate}

\bb{} Stage W4 constructs the sequences $\MM[B_{i,j}]$ for $1\leq j\leq k_i$
and updates the data structures as follows.
\begin{enumerate}
\item 
For each $j$ and each $(h,U)$ in $L[B_{i,j}]$, replace $(h,U)$ by
$(R(h),U)$.
\item 
For each $j$, set $\MM[B_{i,j}]$ to be the sequence of the families
$\CC''_h=\{U~|~(h,U)\in L[B_{i,j}]\}$, where $h$ ranges over those 
integers that are in a pair in $L[B_{i,j}]$.
\item 
Delete $w_i$ and its children from $T$.
\item For each $h\in X$, set $A_2[h]=0$ and $J[h]=\emptyset$.
\end{enumerate}

By Observation~\ref{obs_Shrink} and Lemma \ref{1Reduce}, after the processing of
$w_i$, the invariants hold for $i+1$.

After processing $w_\ell$, we construct $\MM[B_{\ell,0}]$ as follows: 
Replace each pair $(h,U)$ in $L[B_{\ell,0}]$ by $(R(h), U)$, and then 
set $\MM[B_{\ell,0}]$ to be the sequence of the families $\CC''_h=
\{U~|~(h,U)\in L[B_{\ell,0}]\}$, where $h$ ranges over those 
integers that are in a pair in $L[B_{\ell,0}]$.

By the invariants, Observation~\ref{obs_Shrink}, and Lemma \ref{1Reduce}, $\ggg$
satisfies $\MM$ if and only if every block $B$ of $\ggg$ satisfies
$\MM[B]$.  As for the time complexity, we make the following
observations:
\begin{enumerate}
\item 
When processing $w_i$, we create at most $n_i$ new sets all equal
to $\{w_i\}$, where $n_i$ is the maximum number of blocks in a simple graph
with $k_i+1$ vertices. Since $n_i=O(k_i+1)$ and $k_i+1$ does not
exceed the degree of $w_i$ in $\ggg$, the total number of newly
created sets is $O(|\ggg|)$.

\item If a set $U$ does not intersect $\{w_i, \ldots, w_\ell\}$
immediately before the processing of $w_i$, then there is at most one
$w_j\in\{w_i,\ldots,w_\ell\}$ such that some vertices of $U$ are
touched during the processing of $w_j$.
\item If $w_i$ is in $U$ immediately before the processing of $w_i$, then we
	either (1) touch at most $1+|\{v\in U~|~\post(v)\leq
	\post(w_i)\}|$ vertices of $U$ during the processing of $w_i$,
	or (2) touch no vertex of $U$ during the
	processing of each $w_j\in\{w_{i+1},\ldots, w_\ell\}$.
\end{enumerate}
There are at most $q$ unions and $O(\inputsize)$ finds, and at most $|\ggg|$
insertions into each splay tree.  By the above observations, the total time
spent on the union-find data structure is \maintime, that on the splay
trees is $O(\inputsize\log|\ggg|)$, and that on the remaining computation
is $O(\inputsize)$, all within the desired time.
\end{proof}

\section{Case M2 where $\ggg$  is biconnected.}\label{sec:2-con} 
This section assumes that $\ggg$ is biconnected.
Let $\CC_1,\ldots,\CC_q$ be the families in $\MM$.
For each $i\in\{1,\ldots,q\}$, let $\CC_i=\{U_{i,1},\ldots,U_{i,r_i}\}$.

\begin{theorem}\label{thm_m2}
Theorem~\ref{thm_main} holds for Case M2.
\end{theorem}

To prove Theorem~\ref{thm_m2}, we review a decomposition of $\ggg$ in
\S\ref{subsec:spqr}, outline the basic ideas of our CFE algorithm in
\S\ref{subsec:ideas}, detail the algorithm in
\S\ref{subsec:alg}, and analyze it in \S\ref{subsec:ana}.  

\subsection{SPQR decompositions}\label{subsec:spqr}
A {\em planar $st$-graph} $G$ is a directed acyclic plane graph such that
$G$ has exactly one source $s$ and exactly one sink $t$, and both
vertices are on the exterior face. These two vertices are the {\em
poles} of $G$.  A {\em split pair} of $\g$ is either a pair of
adjacent vertices or a pair of vertices whose removal disconnects
the graph obtained from $\g$ by adding the edge $(s,t)$.
A {\em split component} of a split pair $\{u,v\}$ is either an
edge $(u,v)$ or a maximal subgraph $C$ of $\g$ such that $C$ is a
planar $uv$-graph and $\{u,v\}$ is not a split pair of $C$. A split
pair $\{u,v\}$ of $\g$ is {\em maximal} if there is no other split
pair $\{u',v'\}$ in $\g$ such that a split component of $\{u',v'\}$ contains both $u$ and $v$.

The {\em decomposition tree} $T$ of $\g$ is a rooted ordered tree
recursively defined in four cases as follows.  The nodes of $T$ are of
four types $S,P,Q$, and $R$. Each node $\mu$ of $T$ has an associated
planar $st$-graph $\SK(\mu)$, called the {\em skeleton} of
$\mu$. Also, $\mu$ is associated with an edge in the skeleton of the
parent $\phi$ of $\mu$, called the {\em virtual edge} of $\mu$ in
$\SK(\phi)$.

{\it  Case Q}: $\g$ is a single edge from $s$ to $t$.  Then, $T$ is a
Q-node whose skeleton is $\g$.

{\it  Case S}: $\g$ is not biconnected. Let $c_1,\ldots,c_{k-1}$ with
$k\geq 2$ be the cut vertices of $\g$. Since $\g$ is a planar
$st$-graph, each $c_i$ is in exactly two blocks $\g_i$ and $\g_{i+1}$
with $s\in\g_1$ and $t\in\g_k$.  Then, $T$'s root is an S-node $\mu$,
and $\SK(\mu)$ consists of the chain $e_1,\ldots,e_k$, 
where the edge $e_i$ goes from $c_{i-1}$ to $c_i$, $c_0=s$,
and $c_k=t$.

{\it  Case P}: $\{s,t\}$ is a split pair of $\g$ with $k$ 
split components where $k\geq 2$. Then, $T$'s root is a P-node $\mu$, and $\SK(\mu)$
consists of $k$ parallel edges $e_1,\ldots,e_k$ from $s$ to $t$.

{\it  Case R}: Otherwise. Let $\{s_1,t_1\},\ldots,\{s_k,t_k\}$ with
$k\geq 1$ be the maximal split pairs of $\g$.  Let $\g_i$ be the union
of the split components of $\{s_i,t_i\}$. Then, $T$'s root is an
R-node $\mu$, and $\SK(\mu)$ is the simple graph obtained
from $\g$ by replacing each $\g_i$ with an edge $e_i$ from $s_i$ to
$t_i$. Note that adding the edge $(s,t)$ to $\SK(\mu)$ yields a simple triconnected graph.

\begin{figure}[t]
\centerline{\psfig{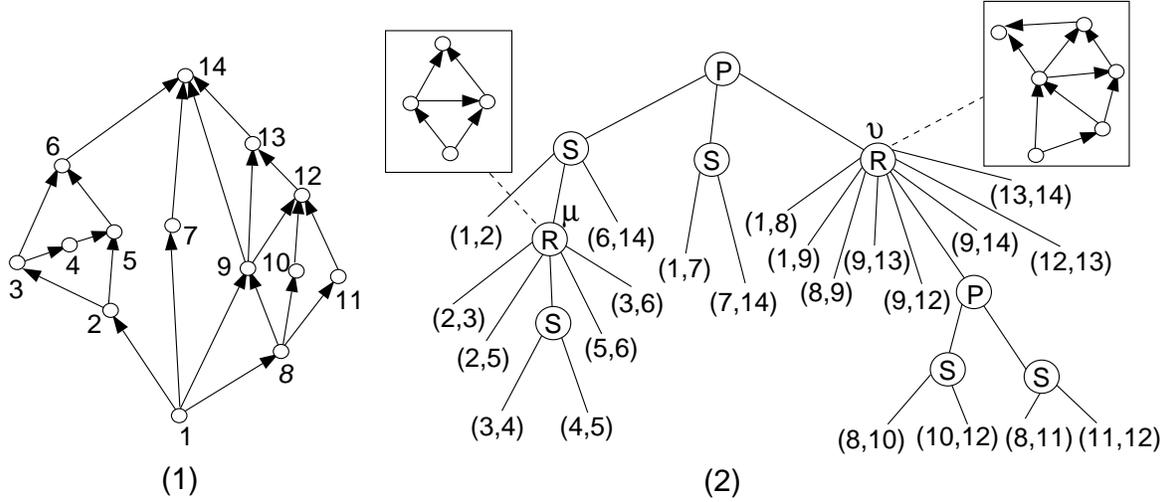}}
\caption{The tree in $($2$)$ is the decomposition tree of the graph in
$($1$)$.}
\label{fig:SPQR}
\end{figure}

Figure \ref{fig:SPQR} illustrates the decomposition tree of $\g$ as
well as the skeletons of $\mu$ and $\nu$.  In the last three cases,
$\mu$ has children $\chi_1,\ldots,\chi_k$ in this order, such that
each $\chi_i$ is the root of the decomposition tree of $\g_i$. The
virtual edge of $\chi_i$ is the edge $e_i$ in $\SK(\mu)$. $\g_i$ is
called the {\em pertinent graph} $\PT(\chi_i)$ of $\chi_i$ as well as
the {\em expansion graph} of $e_i$. Note that $\g$ is the pertinent
graph of $T$'s root.  Also, no child of an S-node is an S-node, and no
child of a P-node is a P-node.

The {\em allocation nodes} of a vertex $v$ of $\g$ are the nodes of
$T$ whose skeleton contains $v$; note that $v$ has at least one
allocation node.

\begin{lemma}[see \cite{DT96}]\label{lemma:linear-space}\

\begin{enumerate}
\item\label{lemma:linear-space-1}
$T$ has $O(|\g|)$ nodes and can be constructed in $O(|\g|)$
time.  The total number of edges of the skeletons stored at the nodes
of $T$ is $O(|\g|)$.
\item 
The pertinent graphs of the children of $\mu$ can only share vertices
of $\SK(\mu)$.
\item If $v$ is in $\SK(\mu)$, then $v$ is also in the
pertinent graph of all ancestors of $\mu$.
\item If $v$ is a pole of $\SK(\mu)$, then $v$ is also in the skeleton
of the parent of $\mu$. If $v$ is in $\SK(\mu)$ but is not a pole of
$\SK(\mu)$, then $v$ is not in the skeleton of any ancestor of $\mu$.
\item The least common ancestor $\mu$ of the allocation nodes of
$v$ itself is an allocation node of $v$, called the {\em proper
allocation node} of $v$. Also, if $v\not\in\{s,t\}$, then $\mu$ is the
only allocation node of $v$ such that $v$ is not a pole of $\SK(\mu)$.
\item If $v\neq s,t$, then the proper allocation node of $v$ is
an R-node or S-node.
\end{enumerate}
\end{lemma}

For each non-S-node $\mu$ in $T$, $\PT(\mu)$ is called a {\em block}
of $\g$ \cite{DT96}, which differs from that in \S\ref{sec:0-con} and
\S\ref{sec:1-con}.  For a block $B=\PT(\mu)$, let $\node(B)=\mu$.  For
an ancestor $\phi$ of $\node(B)$, the {\em representative} of $B$ in
$\SK(\phi)$ is the edge in $\SK(\phi)$ whose expansion graph contains
$B$.

Let $\mu$ be an R-node or P-node in $T$ with children
$\chi_1,\ldots,\chi_b$.  For each $k\in\{1,\ldots,b\}$, let $e_k$ be
the virtual edge of $\chi_k$ in $\SK(\mu)$. If $\chi_k$ is an S-node,
$\PT(\chi_k)$ is a chain consisting of two or more blocks. If $\chi_k$
is an R-node or P-node, $\PT(\chi_k)$ is a single block. For each
$k\in\{1, \ldots, b\}$, we say that the blocks in $\PT(\chi_k)$ are
{\em on edge} $e_k$.  The {\em minor blocks} of $\PT(\mu)$ are the
blocks on $e_1$, \ldots, the blocks on $e_b$.

\subsection{Basic ideas}\label{subsec:ideas}
An {\em $st$-orientation} of a planar graph is an orientation of its
edges together with an embedding such that the resulting digraph is a
planar $st$-graph.
\begin{lemma}[see \cite{BT96b,DT96}]\label{lemma:all-embedding}
If an $n$-vertex simple planar graph has an $st$-orientation, then every
embedding, where $s$ and $t$ are on the exterior face, of this graph
can be obtained from this orientation through
a sequence of $O(n)$ following operations:
\begin{enumerate}
\item 
Flip an R-node's skeleton around its poles.
\item 
Permute a P-node's children 
(and consequently their skeletons with respect to their common poles).
\end{enumerate}
\end{lemma}

Let $\{s,t\}$ be an edge of $\ggg$. Since $\ggg$ is a simple biconnected graph, we
convert $\ggg$ to a planar $st$-graph in $O(n)$ time \cite{ET76a} for
technical convenience.  For the remainder of \S\ref{sec:2-con}, let
$T$ be the decomposition tree of $\ggg$. 

The CFE algorithm processes the nodes of $T$ in a bottom-up manner. 
It first processes the leaf nodes of $T$. When processing a node $\mu$, 
for each $\CC_i$ such that $\PT(\mu)$ is the 
smallest block that intersects every set in $\CC_i$, the 
algorithm looks for an embedding of $\PT(\mu)$ that satisfies 
$\CC_i$.  If this is impossible, the
algorithm outputs ``no'' and stops; otherwise, it continues on to
process the next node of $T$. We note, in passing, that 
Theorem~\ref{thm_m3}(\ref{thm_m3_2}) is used when processing R-nodes. 

Let $\mu$ be a node of $T$. $T_{\mu}$ denotes the subtree of $T$
rooted at $\mu$ and $\depth(\mu)$ denotes the distance from $T$'s root
to $\mu$. We need the following definitions: 
\begin{enumerate}
\item $U_{i,j}$ is {\em contained} in $\PT(\mu)$ if the vertices of
$U_{i,j}$ are all in $\PT(\mu)$; $U_{i,j}$ is {\em strictly contained}
in $\PT(\mu)$ if in addition, no pole of $\PT(\mu)$ is in $U_{i,j}$.

\item 
Let $\settle(U_{i,j})$ be the deepest node $\mu$ in $T$ such that
$U_{i,j}$ is strictly contained in $\PT(\mu)$, if such a node exists.
If no such $\mu$ exists, then $U_{i,j}$ contains a pole of $\ggg$ and
let $\settle(U_{i,j})$ be $T$'s root.

\item A family $\CC_i$ {\em straddles} $\PT(\mu)$ if at least one set
in $\CC_i$ is strictly contained in $\PT(\mu)$, and at least one set
in $\CC_i$ has no vertex in $\PT(\mu)$.

\item 
Let $\settle(\CC_i)$ be the deepest node $\mu$ in $T$ such that for
every $U_{i,j}\in \CC_i$, at least one vertex of $U_{i,j}$ is in
$\PT(\mu)$.

\item Let $\Subset(\mu)=\{ U_{i,j}~|~\settle(U_{i,j})=\mu \}$ and
          $\Fam(\mu)=\{ \CC_i~|~\settle(\CC_i)=\mu\}$.
\item 
If $\mu$ is a P-node or R-node, let
$\EF(\mu)=\Fam(\mu)\cup(\cup_{\chi_k}\Fam(\chi_k))$ and
$\ES(\mu)=\Subset(\mu)\cup(\cup_{\chi_k}\Subset(\chi_k))$, where
$\chi_k$ ranges over all S-children of $\mu$.
\end{enumerate}

In a fixed embedding of a block $B$, the poles of $B$ divide the
boundary of its exterior face into two paths $\side_1(B)$ and
$\side_2(B)$, called the two {\em sides} of $B$.  $U_{i,j}$ is {\em
two-sided} for $B$ if both $\side_1(B)$ and $\side_2(B)$ intersect
$U_{i,j}$. In particular, $U_{i,j}$ is two-sided for $B$ if it
contains a pole of $B$.  $U_{i,j}$ is {\em side-1} (respectively, {\em
side-2}) for $B$ if only $\side_1(B)$ (respectively, $\side_2(B)$)
intersects $U_{i,j}$.  Assume that $B$ is a minor block of $\PT(\mu)$
for some $\mu$.  Let $e_k$ be the representative of $B$ in
$\SK(\mu)$. In a fixed embedding of $\SK(\mu)$, $e_k$ separates two
faces $F$ and $F'$. When embedding $\PT(\mu)$, we can embed
$\side_1(B)$ towards either $F$ or $F'$, referred to as the two {\em
orientations} of $B$ in $\PT(\mu)$.

A family $\CC_i$ is {\em side-0} (respectively, {\em side-1} or {\em
side-2}) {\em exterior-forcing} for $B$ if $\settle(\CC_i)$ is an
ancestor of $\node(B)$ in $T$ and some $U_{i,j} \in \CC_i$ strictly
contained in $B$ is two-sided (respectively, side-1 or side-2) for
$B$.  For $p=0$, 1, 2, define
\begin{enumerate}
\item $\ext_p(B)=$ $\min \{\depth(\settle(\CC_i))~|~\CC_i$, $1\leq i\leq q$,
  is side-$p$ exterior-forcing for $B \}$, if at least one family in $\MM$ 
  is side-$p$ exterior-forcing for $B$;
\item $\ext_p(B)=\infty$ otherwise.
\end{enumerate}

Assume $\ext_p(B)\not=\infty$. Let
$\mu=\node(B),\phi_1,\phi_2,\ldots,\phi_h$ be the path in $T$ from
$\mu$ to $\phi_h$, where $\depth(\phi_h)=\ext_p(B)$.  For each
$\ell\in\{1,\ldots, h-1\}$, the representative of $B$ in
$\SK(\phi_\ell)$ must be an exterior edge in any satisfying embedding
of $\SK(\phi_\ell)$. In addition, if $p=1$ or 2, $\side_p(B)$ must be
embedded towards the exterior face of the embedding of
$\PT(\phi_\ell)$.

Since $(s,t)$ is an edge of $\ggg$, the root $\rho$ of $T$ is 
a P-node and has a child Q-node $\phi$ representing $(s,t)$.
A subtle difference between $\rho$ and each non-root node of $T$ is
that the two sides of $\ggg=\PT(\rho)$ is actually on the same face.
To eliminate this difference, we delete $\phi$ from $T$; afterwards,
if $\rho$ has only one child, we further delete $\rho$ from $T$.
{From} here onwards, $T$ denotes this modified tree.

\subsection{The CFE algorithm}\label{subsec:alg}
The CFE algorithm processes $T$ from bottom up.  A {\em ready} node
$\mu$ of $T$ is either (1) a leaf node or (2) a P-node or R-node such
that the non-S-children of $\mu$ and the children of every S-child of
$\mu$ all have been processed.  The CFE algorithm processes the ready
nodes of $T$ in an arbitrary order.  An S-node is processed when its
parent is processed.  We detail how to process $\mu$ as follows.

For the case where $\mu$ is a leaf node of $T$, note that $\PT(\mu)$
is a single edge of $\ggg$. Since no $U_{i,j}$ is strictly contained in
$\PT(\mu)$, $\Subset(\mu)=\emptyset$. Also, each $\CC_i\in \Fam(\mu)$
is satisfied by every embedding of $\ggg$. Therefore, we simply
set $\ext_p(\PT(\mu))=\infty$ for $p=0,1,2$.

We next consider the case where $\mu$ is a non-leaf ready node.
Before $\mu$ is processed, an embedding of every minor block of
$\PT(\mu)$ is already fixed, except for a possible flip around its
poles. Moreover, for each minor block $B$ of $\PT(\mu)$ and each
$p\in\{0,1,2\}$, $\ext_p(B)$ is known.  When processing $\mu$, the CFE
algorithm checks whether some embedding ${\EE}_{\mu}$ of $\PT(\mu)$
satisfies the following two conditions:

\begin{enumerate}
\item ${\EE}_{\mu}$ satisfies every $\CC_i$ in $\EF(\mu)$.

\item For each $\CC_i$ straddling $\PT(\mu)$ and each $U_{i,j}\in
\CC_i$ strictly contained in $\PT(\mu)$, at least one vertex of
$U_{i,j}$ is embedded on the exterior face of ${\EE}_{\mu}$. ({\it 
Remark.} This ensures the existence of an embedding of
$\PT(\settle(\CC_i))$ satisfying $\CC_i$ later.)
\end{enumerate}
If no such ${\EE}_{\mu}$ exists, then $\ggg$ cannot satisfy $\MM$ and
the CFE algorithm outputs ``no'' and stops.  Otherwise, it finds such
an ${\EE}_{\mu}$ and fixes it except for a possible flip around its
poles.  It also computes $\ext_p(\PT(\mu))$ for $p=0$, 1, 2.

To detail how to process $\mu$, we classify the sets $U_{i,j}$ that intersect
$\PT(\mu)$ into four types and define a set $\Image(U_{i,j},\mu)$ for
each type as follows.

{\em Type 1}: $U_{i,j}$ contains at least one pole of $\SK(\mu)$.
Then, $\settle(U_{i,j})$ is an ancestor of $\mu$. Let
$\Image(U_{i,j},\mu)=\{v\in U_{i,j}~|~v$ is a vertex in $\SK(\mu)\}$.

{\em Type 2}: $U_{i,j}$ contains at least one vertex but no pole of
$\SK(\mu)$. Then, $\settle(U_{i,j})=\mu$.  Let $\Image(U_{i,j},\mu)$
as in the case of type 1.

{\em Type 3}: $U_{i,j}$ is strictly contained in $\PT(\chi)$ for some
S-node child $\chi$ of $\mu$ and $U_{i,j}$ contains at least one vertex in
$\SK(\chi)$. Then, $\settle(U_{i,j})=\chi$. Let $\Image(U_{i,j},\mu)$
consist of the virtual edge of $\chi$ in $\SK(\mu)$.

{\em Type 4}: $U_{i,j}$ is strictly contained in a minor block $B$ of
$\PT(\mu)$. Then, $\settle(U_{i,j})$ is $\node(B)$ or its
descendent. Let $\Image(U_{i,j},\mu)$ consist of the representative of
$B$ in $\SK(\mu)$.

Each element of $\Image(U_{i,j},\mu)$ is called an {\em image} of
$U_{i,j}$ in $\SK(\mu)$.  The remainder of \S\ref{subsec:alg} details
how to process $\mu$.

\subsubsection{Processing an S-child of $\mu$}
When processing $\mu$, for each S-child $\chi$ of $\mu$, we need to
find an embedding of $\PT(\chi)$ satisfying certain conditions.  We
call this process the {\em S-procedure} and describe it below.

Let $\chi$ be an S-child of $\mu$. Then, $\SK(\chi)$ is a path.  Let
$e_1$, \ldots, $e_b$ be the edges in $\SK(\chi)$.  For each
$k\in\{1,\ldots,b\}$, let $B_k$ be the expansion graph of $e_k$.
Before the S-procedure is called on $\chi$, the following requirements
are met:
\begin{enumerate}
\item 
For each $k\in\{1,\ldots,b\}$, an embedding of $B_k$ has been fixed,
except for a possible flip around its poles.
\item\label{Require}
For some integers $k\in\{1,\ldots,b\}$ and $p\in\{1,2\}$,
$\side_p(B_k)$ is required to face either the left or the right side
of $\SK(\chi)$.
\end{enumerate}
Our only choice for embedding $\PT(\chi)$ is to flip $B_1$, \ldots,
$B_b$ around their poles. We need to check whether for some
combination of flippings of $B_1$, \ldots, $B_b$, (1) the resulting
embedding satisfies every $\CC_i\in\Fam(\chi)$ and (2) the second
requirement above is met.

The S-procedure consists of the following five stages:

\bb{} Stage S1 constructs an auxiliary graph $D=(V_D,E_D)$ with $V_D=\{
k_p~|~1 \leq k \leq b,~p=1,2 \}$ as follows.  For each $\CC_i \in
\Fam(\chi)$, insert an arbitrary path $P_i$ into $D$ to connect all $k_p\in V_D$
such that for some type-4 $U_{i,j} \in \CC_i$, (a)
$\Image(U_{i,j},\chi)=\{e_k\}$ and (b) $U_{i,j}$ is side-$p$ for
$B_k$.  To avoid confusion, we call the elements of $V_D$ {\em
points}, and the connected components of $D$ {\em clusters}.  Those
points $k_p\in V_D$ such that $\side_p(B_k)$ is required to face the
left side of $\SK(\chi)$ are called {\em $L$-points}. {\em R-points}
are defined similarly.  Note that for each cluster $C$ of $D$, all
$\side_p(B_k)$ where $k_p$ ranges over all the points in $C$ must be
embedded toward the same side of $\SK(\chi)$.  Also, each type-3
$U_{i,j}$ in $\CC_i$ contains a vertex in $\SK(\chi)$ which is on both
sides of $\SK(\chi)$. For this reason, such sets were not considered
when constructing $D$.

\bb{} Stage S2 checks whether there is a cluster of $D$ containing both an
$L$-point and an R-point. If such a cluster exists, then S2 outputs
``no" and stops. Suppose that no such cluster exists. 
If a cluster $C$ contains an $L$-point (respectively, $R$-point), 
we call $C$ an {\em $L$-cluster} (respectively, {\em $R$-cluster}).

\bb{} Stage S3 constructs another auxiliary graph $RD=(V_{RD}, E_{RD})$ from
$D$ as follows. The vertices of $RD$ are the clusters of $D$.  For
each $k\in\{1, \ldots, b\}$, there is an edge $\{C_1, C_2\}$ in $RD$,
where $C_1$ (respectively, $C_2$) is the cluster of $D$ containing
point $k_1$ (respectively, $k_2$). Note that $RD$ may have self-loops.

\bb{} Stage S4 checks whether $RD$ is bipartite. If it is not, then S4
outputs ``no'' and stops. Otherwise, for each connected component $K$
of $RD$, the clusters in $K$ can be uniquely partitioned into two
independent subsets $V_{K,1}$ and $V_{K,2}$ of clusters. If $V_{K,1}$
or $V_{K,2}$ contains both an $L$-cluster and an R-cluster, S4 outputs
``no" and stops. Otherwise, $V_{RD}$ can be partitioned into two
independent subsets $V_{RD}^L$ and $V_{RD}^R$ of clusters such that
all $L$-clusters are in $V_{RD}^L$ and all R-clusters are in
$V_{RD}^R$.  Let $V_D^L=\{ k_p~|~k_p$ is in a cluster in $V_{RD}^L\}$
and $V_D^R=\{ k_p~|~k_p$ is in a cluster in $V_{RD}^R\}$.

\bb{} Stage S5 embeds $\side_p(B_k)$ toward the left side of $\SK(\chi)$ for
each $k_p\in V_D^L$.

\begin{figure}[t]
\centerline{\psfig{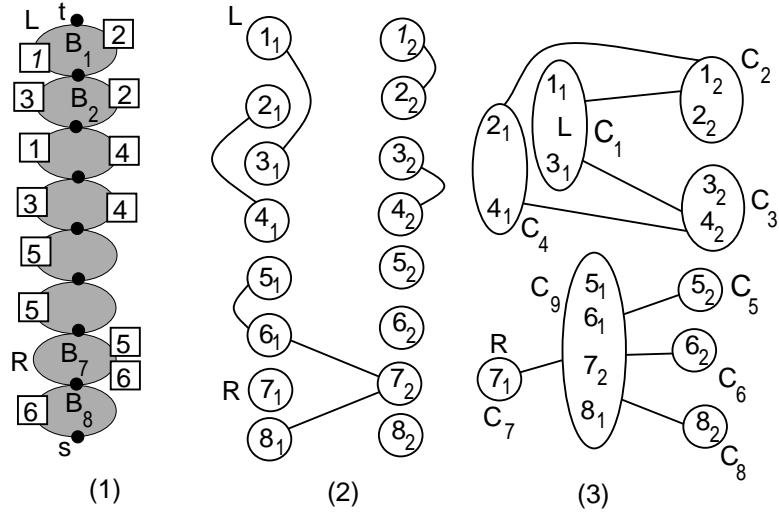}}
\caption{The graph in $($1$)$ is $\PT(\chi)$ for an S-node $\chi$,
the graph in $($2$)$ is $D$, and that in $($3$)$ is $RD$.}
\label{fig-S}
\end{figure}

\begin{example}\label{ex-S}\rm
In Figure \ref{fig-S}, $\PT(\chi)$ has 8 blocks $B_1,\ldots,B_8$. The
left side of each $B_k$ is $\side_1(B_k)$.  Also,
$\Fam(\chi)=\{\CC_1,\ldots,\CC_6\}$.  An integer $i$ in a small square
on $\side_p(B_k)$ for $p=1$ or 2 indicates that $k_p$ is on $P_i$. For
example, the points on $P_5$ are $5_1$, $6_1$, and $7_2$. The letter
$L$ is marked on $\side_1(B_1)$, indicating that $\side_1(B_1)$ must
face left. The letter $R$ is marked on $\side_1(B_7)$, indicating that
$\side_1(B_7)$ must face right.  $D$ is shown in
Figure~\ref{fig-S}(2). $1_1$ is an $L$-point while $7_1$ is an
R-point.  $RD$ is shown in Figure~\ref{fig-S}(3). $C_1$ is an
$L$-cluster and $C_7$ is an R-cluster.  $RD$ is bipartite and $V_{RD}$
can be partitioned into $V_{RD}^L=\{C_1,C_4,C_9\}$ and
$V_{RD}^R=\{C_2,C_3,C_5,C_6,C_7,C_8\}$.  Thus
$V_D^L=\{1_1,2_1,3_1,4_1,5_1,6_1,7_2,8_1\}$ and
$V_D^R=\{1_2,2_2,3_2,4_2, 5_2,6_2,7_1,8_2\}$. Flipping $B_7$ in
Figure~\ref{fig-S}(1) gives a satisfying embedding of $\PT(\chi)$.  If
$8_2$ were also on $P_5$, there would be an edge $\{7_2,8_2\}$ in $D$,
which would cause $C_9$ and $C_8$ to be merged in $RD$ with a
self-loop attached to it. In that case, $RD$ would not be bipartite
and the CFE algorithm would output ``no''.
\end{example}

\subsubsection{$\mu$ is an R-node}\label{subsec:Rcase}

In this case, adding the edge $(s,t)$ to $\SK(\mu)$ yields a simple triconnected graph. 
Thus, the unique embedding of $\SK(\mu)$ with both $s$ and $t$ on the exterior face is 
$\SK(\mu)$ itself. Let $\chi_1$,\ldots, $\chi_b$ be the children of $\mu$
in $T$. For each $k\in\{1, \ldots, b\}$, let $B_{k,1}$,\ldots,
$B_{k,s_k}$ be the minor blocks of $\PT(\mu)$ in $\PT(\chi_k)$.  Note
that $s_k=1$ when $\chi_k$ is an R-node or P-node.  To process $\mu$,
the CFE algorithm proceeds in five stages:

\bb{} Stage R1 first computes $\CC'_i=\{ \Image(U_{i,j},\mu)~|~U_{i,j} \in \CC_i
\}$ for every $\CC_i \in \Fam(\mu)$. Let $\MM'(\mu)$ be the
sequence of all $\CC'_i$ with $\CC_i \in \Fam(\mu)$. Then R1 calls
Theorem~\ref{thm_m3}(\ref{thm_m3_2}) to solve the CFE problem on input
$\SK(\mu)$ and $\MM'(\mu)$. If the output is ``no", R1 outputs ``no''
and stops. Otherwise, for each $\CC'_i$ in $\MM'(\mu)$, there is a
face $F_i$ in $\SK(\mu)$ whose boundary
intersects each $\Image(U_{i,j},\mu) \in \CC'_i$.  Note that $F_i$
must be unique or else $\settle(\CC_i)$ would be a descendent of
$\mu$, contradicting the fact $\CC_i \in \Fam(\mu)$.

\bb{} Stage R2 computes the minor block $B_{k,l}$ of $\PT(\mu)$ strictly
containing $U_{i,j}$ for each $\CC_i \in \Fam(\mu)$ and each type-4
$U_{i,j} \in \CC_i$.  If $U_{i,j}$ is two-sided for
$B_{k,l}$, either side of $B_{k,l}$ may be embedded toward the face
$F_i$; otherwise, for some $p \in \{1,2\}$, $U_{i,j}$ is side-$p$ for
$B_{k,l}$ and it requires that $\side_p(B_{k,l})$ be embedded towards
$F_i$.

\bb{} Stage R3 makes sure that for every $\CC_i$ straddling
$\PT(\mu)$ and for every $U_{i,j} \in \CC_i$ strictly contained in $\PT(\mu)$,
a vertex in $U_{i,j}$ is embedded on the exterior face of $\PT(\mu)$.
This is done by checking whether the following statements are all false.
\begin{enumerate}
\item\label{bad-case-a}
There are an exterior edge $e_k$ of $\SK(\mu)$ and a minor block
$B_{k,l}$ of $\PT(\mu)$ on $e_k$ with $\max_{p \in \{1,2\}}
\ext_p(B_{k,l}) < \depth(\mu)$; thus, both
$\side_1(B_{k,l})$ and $\side_2(B_{k,l})$ must be embedded towards the
exterior face of $\SK(\mu)$.
\item\label{bad-case-b}
There are an interior edge $e_k$ of $\SK(\mu)$ and a minor block
$B_{k,l}$ of $\PT(\mu)$ on $e_k$ with $\min_{p\in\{0,1,2\}}
\ext_p(B_{k,l}) < \depth(\mu)$; thus, at least one of
$\side_1(B_{k,l})$ and $\side_2(B_{k,l})$ must be embedded towards the
exterior face of $\SK(\mu)$.
\item\label{bad-case-c} There is a $U_{i,j} \in \Subset(\mu)$ with
$\depth(\settle(\CC_i)) < \depth (\mu)$ (i.e., $\CC_i$ straddles
$\PT(\mu)$) and neither side of $\SK(\mu)$ contains
an image in $\Image(U_{i,j},\mu)$.
\item\label{bad-case-d} There are an S-child $\chi_k$ of $\mu$ and
a $U_{i,j} \in \Subset(\chi_k)$ such that $\depth(\settle(\CC_i)) <
 \depth(\mu)$ and the virtual edge $e_k$ of $\chi_k$ is an interior edge
in $\SK(\mu)$.
\end{enumerate}
If at least one statement above holds, R3 outputs ``no'' and stops.
Otherwise, for each minor block $B_{k,l}$ of
$\PT(\mu)$ such that $\ext_p(B_{k,l}) < \depth(\mu)$ for some $p\in\{1,2\}$,
it requires that $\side_p(B_{k,l})$ be embedded towards the exterior
face of $\SK(\mu)$. Note that since the above \ref{bad-case-b} is false, the
representative $e_k$ of $B_{k,l}$ in $\SK(\mu)$ must be
an exterior edge of $\SK(\mu)$.

\bb{} Stage R4 first checks whether for some minor block $B_{k,l}$ of
$\PT(\mu)$, the orientation requirements imposed on $B_{k,l}$ in Stage
R2 or R3 are in conflict. If they are, R4 outputs ``no'' and stops.
Otherwise, for each R-child or P-child $\chi_k$ of $\mu$, the minor
block $\PT(\chi_k)$ can be oriented according to the requirements
imposed on it, or arbitrarily if no requirement was imposed on
it. Afterwards, for each S-child $\chi_k$ of $\mu$, it calls the
S-procedure on input $\chi_k$ together with the orientation
requirements that were imposed on the minor blocks in $\PT(\chi_k)$ in
Stage R2 or R3.  If the S-procedure on a $\chi_k$ outputs ``no",
R4 outputs ``no'' and stops because $\PT(\chi_k)$ cannot be
successfully embedded; otherwise, it has found a satisfying embedding
of $\PT(\mu)$.

\bb{} Stage R5 computes $\ext_p(\PT(\mu))$ for $p=0$, 1, 2 as follows.  Let
$\ES'(\mu)=\{ U_{i,j} \in \ES(\mu)~|$ $\depth(\settle(\CC_i)) <
\depth(\mu)\}$; i.e., $\ES'(\mu)$ consists of all $U_{i,j}\in \ES(\mu)$
such that $\CC_i$ straddles $\PT(\mu)$.  Partition $\ES'(\mu)$ into
$A_0$, $A_1$, $A_2$ where $A_0$ (respectively, $A_1$ or $A_2$)
consists of all $U_{i,j} \in \ES'(\mu)$ such that $U_{i,j}$ is
two-sided (respectively, side-1 or side-2) for $\PT(\mu)$. For
$i\in\{1,2\}$, let $\beta_i = \min_{p,B_{k,l}} \ext_p(B_{k,l})$ where
$p$ ranges over all integers in $\{0,1,2\}$ and $B_{k,l}$ ranges over
all minor blocks on an edge of $\side_i(\SK(\mu))$. Then, set
\begin{eqnarray*}
\label{eqn:A0}
\ext_0(\PT(\mu)) & = & \min_{U_{i,j} \in A_0}\depth(\settle(\CC_i));
\\ \label{eqn:A1} 
\ext_1(\PT(\mu)) & = & \min \{\beta_1, \min_{U_{i,j} \in A_1}
\depth(\settle(\CC_i))\};
\\
\label{eqn:A2}
\ext_2(\PT(\mu)) & = & \min \{\beta_2,\min_{U_{i,j} \in A_2}
\depth(\settle(\CC_i))\}.
\end{eqnarray*}
This completes the processing of $\mu$.

\begin{figure}[t]
\centerline{\psfig{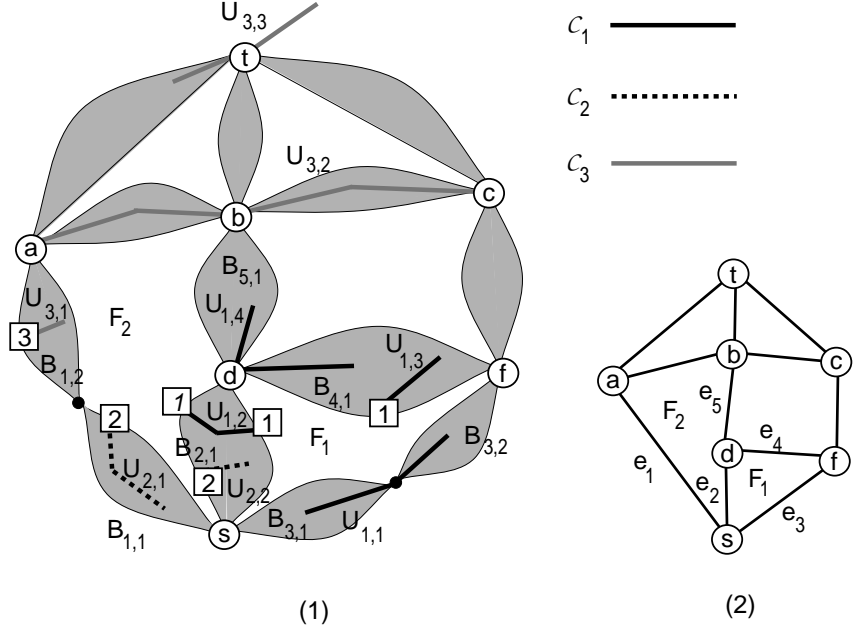}}
\caption{The graph in $($1$)$ is $\PT(\mu)$ for an R-node $\mu$, and
the graph in $($2$)$ is $\SK(\mu)$.}
\label{fig-R}
\end{figure}

\begin{example}\label{ex-R}\rm
In Figure~\ref{fig-R}, the circles denote the vertices in $\SK(\mu)$,
where $s$ and $t$ are the poles of $\PT(\mu)$. An integer $i$ in a
small square at a side of a block $B_{k,l}$ indicates that a set in
$\CC_i$ has a vertex on that side of $B_{k,l}$.  Also,
$\Fam(\mu)=\{\CC_1,\CC_2\}$.
$\CC_1=\{U_{1,1},\ldots,U_{1,4}\}$. $U_{1,1}$ is of type 3 and
$\Image(U_{1,1},\mu)=\{e_3\}$. $U_{1,2}$ and $U_{1,3}$ are of type 4,
$\Image(U_{1,2},\mu)=\{e_2\}$, and $\Image(U_{1,3},\mu)
=\{e_4\}$. $U_{1,2}$ is two-sided for $B_{2,1}$. $U_{1,4}$ is of type 2
and $\Image(U_{1,4},\mu)=\{d\}$. $\CC_2$ consists of $U_{2,1}$ and
$U_{2,2}$, which are of type 4.  $\Image(U_{2,1},\mu)=\{e_1\}$ and
$\Image(U_{2,2},\mu)=\{e_2\}$.  $\CC_3$ is the only family straddling
$\PT(\mu)$.  $U_{3,1}$, $U_{3,2}$, and $U_{3,3}$ are the sets in $\CC_3$
that intersect $\PT(\mu)$; the other sets in $\CC_3$ are not shown 
in this figure.  $U_{3,1}$ is of type 4 and
$\Image(U_{3,1},\mu)=\{e_1\}$.  $U_{3,2}$ is of type 2 and is two-sided
for $\PT(\mu)$; $\Image(U_{3,2},\mu)=\{a,b,c\}$. Since $U_{3,3}$ is
not strictly contained in $\PT(\mu)$, it is not tested during the
processing of $\mu$. Note that $\PT(\mu)$ has a satisfying embedding
as shown.  For $i=1, 2$, the boundary of $F_i$ intersects each set in
$\CC_i$.  The exterior face of $\PT(\mu)$ contains an image of every
set in $\CC_3$ strictly contained in $\PT(\mu)$.  The side of
$B_{4,1}$ on which $1$ is marked must be embedded toward $F_1$.  In
contrast, whichever side of $B_{2,1}$ is embedded toward $F_1$, the
boundary of $F_1$ intersects $U_{1,2}$.  In the embedding of
$\PT(\mu)$, $\CC_3$ is side-1 (respectively, side-0) exterior-forcing
for $\PT(\mu)$ because of $U_{3,1}$ (respectively, $U_{3,2}$).
\end{example}

\subsubsection{$\mu$ is a P-node}\label{subsec:Pcase}
In this case, $\SK(\mu)$ consists of parallel edges
$e_1,e_2,\ldots,e_b$ between its two poles with $b\geq 2$.  Let
$\chi_1$,\ldots, $\chi_b$ be the children of $\mu$ in $T$.  For each
$k\in\{1, \ldots, b\}$, let $B_{k,1}$,\ldots, $B_{k,s_k}$ be the minor
blocks of $\PT(\mu)$ in $\PT(\chi_k)$.  When embedding $\SK(\mu)$,
edges $e_1$ through $e_b$ can be embedded in any order.  The CFE
algorithm first finds a proper embedding of $\SK(\mu)$ in three
stages:

\bb{} Stage P1 constructs an auxiliary graph $\hh=(V_{\hh}, E_{\hh})$ with
$V_{\hh}=\{e_1,\ldots,e_b\}$ by performing the following steps in turn for
every $\CC_i \in \Fam(\mu)$:
\begin{enumerate}
\item Compute $S_i=\cup_{U_{i,j}}\Image(U_{i,j}, \mu)$, where
$U_{i,j}$ ranges over all type-3 or type-4 sets in $\CC_i$.  Let $m_i$
be the number of edges in $S_i$.  Then, $m_i \geq 2$; otherwise
$\CC_i$ would be in $\Fam(\chi_k)$ for some $k\in\{1,\ldots,b\}$.
\item If $m_i\geq 3$, then output ``no" and stop since $\PT(\mu)$ does
not satisfy $\CC_i$.
\item Insert edge $\{e_k,e_{k'}\}$ to $\hh$, where $e_k$ and $e_{k'}$
	are the two edges in $S_i$.
\end{enumerate}
Note that for each $\CC_i \in \Fam(\mu)$, no set in $\CC_i$ is of type 2,
and each type-1 set in $\CC_i$ contains a pole of $\PT(\mu)$, which is
on every face of all embeddings of $\SK(\mu)$. For this reason,
neither type-1 nor type-2 set in $\CC_i$ is considered in the
construction of $\hh$.

\bb{} Stage P2 checks whether both statements below are false in order to
ensure that for every $\CC_i$ straddling $\PT(\mu)$ and every $U_{i,j}
\in \CC_i$ strictly contained in $\PT(\mu)$, a vertex in $U_{i,j}$ is
embedded on the exterior face of $\PT(\mu)$.
\begin{enumerate}
\item\label{bad-case-a-for-P}
There is a minor block $B_{k,l}$ of $\PT(\mu)$ with $\max_{p\in\{
1,2\}} \ext_p(B_{k,l}) < \depth (\mu)$.
\item\label{bad-case-b-for-P} There are at least three edges $e_k$ in
$\SK(\mu)$ such that (1) there is a minor block $B_{k,l}$ on $e_k$
with $\min_{p\in \{ 0,1,2\}} \ext_p(B_{k,l}) < \depth(\mu)$; or (2)
$\chi_k$ is an S-node and there exists $U_{i,j}$ in $\Subset(\chi_k)$
with $\depth(\settle(\CC_i)) < \depth(\mu)$.
\end{enumerate}
If Statement 1 or 2 holds, P2 outputs ``no" and stops. Otherwise, it
marks each $e_k\in V_{\hh}$ for which Statement
\ref{bad-case-b-for-P}(a) or \ref{bad-case-b-for-P}(b) holds. Note
that at most two $e_k\in V_{\hh}$ are marked, and each marked $e_k\in
V_{\hh}$ must be an exterior edge in any satisfying embedding of
$\SK(\mu)$.

\bb{} Stage P3 outputs ``no" and stops if an $e_k\in V_{\hh}$ has degree at
least three in $\hh$ or a marked $e_k\in V_{\hh}$ has degree 2 in
$\hh$.  Otherwise, P3 finds and fixes an embedding of $\SK(\mu)$ where
(1) each marked $e_k\in V_{\hh}$ is in the exterior face and (2) for
every $\{e_k,e_{k'}\}\in E_{\hh}$, $e_k$ and $e_{k'}$ form the
boundary of a face.  For each $\CC_i\in\Fam(\mu)$, let $F_i$ be the
face in the fixed embedding of $\SK(\mu)$ whose boundary is formed by
the two edges in $S_i$.  Note that for each $U_{i,j}\in\CC_i$, the
boundary of $F_i$ intersects $\Image(U_{i,j},\mu)$.

Next, the CFE algorithm tries to embed $\PT(\mu)$ based on the
embedding of $\SK(\mu)$ fixed in Stage P3 through the same stages as
Stages R2 through R5 in \S\ref{subsec:Rcase} except that in the stage
corresponding to R5, $A_0=\emptyset$ and the algorithm sets
$\ext_0(\PT(\mu))=\infty$.  This completes the processing of $\mu$.

\begin{figure}[t]
\centerline{\psfig{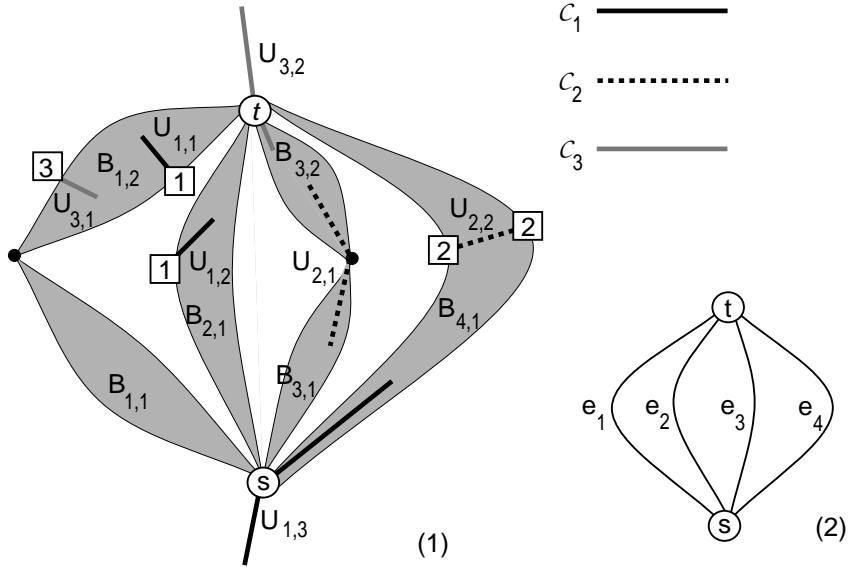}}
\caption{The graph in $($1$)$ is $\PT(\mu)$ for a P-node $\mu$, and
that in $($2$)$ is $\SK(\mu)$.}
\label{fig-P}
\end{figure}

\begin{example}\label{ex-P}\rm
In Figure \ref{fig-P}, $\Fam(\mu)=\{\CC_1,\CC_2\}$.
$\CC_1=\{U_{1,1},U_{1,2},U_{1,3}\}$.  Both $U_{1,1}$ and $U_{1,2}$ are
of type 4; $\Image(U_{1,1}, \mu)=\{e_1\}$ and $\Image(U_{1,2},
\mu)=\{e_2\}$. $U_{1,3}$ is of type 1 and needs not be tested during the
processing of $\mu$. $\CC_2= \{U_{2,1}, U_{2,2}\}$. $U_{2,1}$ is of type
3 and $\Image(U_{2,1}, \mu)=\{e_3\}$.  $U_{2,2}$ is of type 4 and
$\Image(U_{2,2}, \mu)=\{e_4\}$.  $\CC_3$ is the only family straddling
$\PT(\mu)$.  $\{U_{3,1}$ and $U_{3,2}\}$ are the sets in $\CC_3$ that 
intersect $\PT(\mu)$; the other sets in $\CC_3$ are not shown in this figure. 
Since $U_{3,2}$ contains the
pole $t$ of $\PT(\mu)$, it is not tested during the processing of
$\mu$.  $U_{3,1}$ is of type 4 and $\Image(U_{3,1}, \mu)=\{e_1\}$.
$V_{\hh}=\{e_1,e_2,e_3,e_4\}$ and $E_{\hh}=\{\{e_1,e_2\},
\{e_3,e_4\}\}$.  Only $e_1$ is marked in graph
$\hh$. Figure~\ref{fig-P}(2) shows an embedding of $\SK(\mu)$ that
might be found and fixed in Stage P3. This embedding of $\SK(\mu)$
results in a satisfying embedding of $\PT(\mu)$ as shown.  If either
$\CC_1$ had another set strictly contained in block $B_{4,1}$ or
$\CC_3$ had another set strictly contained in $B_{2,1}$, then
$\PT(\mu)$ has no satisfying embedding.
\end{example}

This completes the description of the CFE algorithm.  Its correctness
follows from the above discussion and Fact \ref{lemma:all-embedding}.

\subsection{Implementation and analysis}\label{subsec:ana}
We implement the CFE algorithm as follows.  The nodes of $T$ are
identified by their pre-order numbers. At each node $\mu \in T$, we
store $\depth(\mu)$ and the pre-order number of the largest node in
$T_{\mu}$.  Let $\chi_1,\ldots,\chi_b$ be the children of $\mu$.  The
nodes in $T_{\chi_1},\ldots,T_{\chi_b}$ form an ordered partition of
the nodes in $T_{\mu}-\{ \mu\}$. For a node $\nu$, we can check
whether $\nu$ is in $T_{\mu}$ in $O(1)$ time. If $\nu \in T_{\mu}$, we
can find the subtree $T_{\chi_k}$ containing $\nu$ in $O(\log |\ggg|)$
time, by binary searching the children of $\mu$. We equip $T$ with a
data structure which can be constructed in linear time and outputs a
least common ancestor query in $O(1)$ time \cite{HaTa84,SV88}.

We also store $\SK(\mu)$ at $\mu$. Each $\mu$ has a pointer to its
virtual edge in its parent's skeleton. For each non-pole vertex of
$\SK(\mu)$, we mark $\mu$ as its proper allocation node. This takes
$O(|\ggg|)$ total time by
Fact~\ref{lemma:linear-space}(\ref{lemma:linear-space-1}). Each edge
$e$ of $\ggg$ has a pointer to the leaf node in $T$ that represents
$e$.

\begin{lemma}\label{lemma:pre} 
Given $\ggg$, $\MM$, and $T$, we can compute $\Fam(\mu)$,
$\Subset(\mu)$, $\settle(\CC_i)$, and $\settle(U_{i,j})$ for all nodes
$\mu$ of $T$, all $\CC_i$ in $\MM$, and all $U_{i,j}$ in $\CC_i$ in
$O(\inputsize)$ total time.
\end{lemma}
\begin{proof}
For each vertex $v$ of $\ggg$, let $\low(v)$ be the deepest allocation
node of $v$ in $T$. In $O(|\ggg|)$ time, we can compute $\low(v)$ for
all vertices $v$ of $\ggg$. For a set $U_{i,j} \in \CC_i$, if a pole of 
$\ggg$ is in $U_{i,j}$, then $\settle(U_{i,j})$ is the root of $T$; otherwise, 
$\settle(U_{i,j})$ is the least common ancestor of all $\low(v)$ with
$v \in U_{i,j}$. So, $\settle(U_{i,j})$ can be computed in $O(|U_{i,j}|)$ time. 
Let $\low(U_{i,j})$ be the deepest one among all $\low(v)$ with $v\in
U_{i,j}$.  We can compute $\low(U_{i,j})$ in $O(|U_{i,j}|)$ time.
Since $\settle(\CC_i)$ is the least common ancestor of all
$\low(U_{i,j})$ with $U_{i,j} \in \CC_i$, it can be computed in
$O(|\CC_i|)$ time. Thus, in $O(\inputsize)$ total time, we can compute
$\settle(U_{i,j})$ and $\settle(\CC_i)$ for all $\CC_i$ in $\MM$ and
all $U_{i,j}$ in $\CC_i$. Afterwards, in $O(\inputsize)$ total time, we
can compute $\Fam(\mu)$ and $\Subset(\mu)$ for all nodes $\mu$ of $T$.
\end{proof}

After processing $\mu$, the CFE algorithm records the following
information:
\begin{enumerate}
\item the embedding of $\SK(\mu)$;
\item $\ext_p(\PT(\mu))$ for $p=0$, 1, and 2;
\item the edges and vertices on $\side_1(\SK(\mu))$ and $\side_2(\SK(\mu))$,
	respectively;
\item 
an integer $p=0$, 1 or 2, for each $U_{i,j} \in \ES(\mu)$, indicating
whether $U_{i,j}$ is two-sided, side-1, or side-2 for $\PT(\mu)$,
respectively.
\end{enumerate}

The CFE algorithm processes a P-node or R-node $\mu$ with the five
operations below.

Operation 1 uses $O(|U_{i,j}|+\log |\ggg|)$ time to determine the type
of a given $U_{i,j}$ in $\EF(\mu)$ and finds $\Image(U_{i,j},\mu)$ as
follows.  Let $\nu=\settle(U_{i,j})$.

{\em Case 1}: $\depth(\nu)\leq \depth(\mu)$.  Then, $U_{i,j}$ is
of type 1 or 2 for $\PT(\mu)$. $U_{i,j}$ is of type 1 if and only if it
contains a pole of $\PT(\mu)$. Also, $\Image(U_{i,j},\mu)$ consists of
all $v \in U_{i,j}$ which are also in $\SK(\mu)$. Note that
$v\in\SK(\mu)$ if and only if $\mu$ is the proper allocation node of
$v$ or $v$ is a pole of $\PT(\mu)$.

{\em Case 2}: $\depth(\nu) = \depth(\mu)+1$ and $\nu$ is an
S-node.  Then, $U_{i,j}$ is of type 3 for $\PT(\mu)$. Also,
$\Image(U_{i,j},\mu)$ consists of the virtual edge of $\nu$ in
$\SK(\mu)$.

{\em Case 3}: otherwise. Then, $U_{i,j}$ is of type-4 for
$\PT(\mu)$. Also, $\Image(U_{i,j},\mu)$ is the virtual edge of
$\chi_k$ in $\SK(\mu)$, where $\chi_k$ is the child of $\mu$ such that
$\nu$ is in the subtree $T_{\chi_k}$.

Operation 2 checks in $O(|U_{i,j}|)$ time whether a given $U_{i,j} \in
\ES(\mu)$ has a vertex on either side of $\PT(\mu)$ after an embedding
of $\PT(\mu)$ is fixed.  If $U_{i,j}\in\Subset(\mu)$, we check whether
a vertex in $\Image(U_{i,j},\mu)$ is on either side of $\SK(\mu)$. If
$U_{i,j} \in \Subset(\chi_k)$ for an S-child $\chi_k$ of $\mu$, we
check whether the virtual edge $e_k$ of $\chi_k$ is on either side of
$\SK(\mu)$.

Operation 3 uses $O(1)$ time to check whether a given $U_{i,j} \in
\ES(\mu)$ is in $\ES'(\mu)$ by checking whether
$\depth(\settle(\CC_i)) < \depth(\mu)$.

Operation 4 checks whether a given $U_{i,j}$ is strictly contained in
$\PT(\mu)$ and if so, further computes the minor block $B$ of
$\PT(\mu)$ strictly containing $U_{i,j}$ in $O(|U_{i,j}|+\log |\ggg|)$
total time.  For the first task, we check whether (1) $\nu=\settle(U_{i,j})$
is a descendent of $\mu$, or (2) $\nu=\mu$ and $U_{i,j}$
contains no pole of $\PT(\mu)$.  For the second task, we first find
the child $\chi_k$ of $\mu$ such that $T_{\chi_k}$ contains $\nu$.  If
$\chi_k$ is not an S-node, $\PT(\chi_k)$ is $B$; otherwise, $B$ is
$\PT(\eta)$ where $\eta$ is the child of $\chi_k$ such that $T_{\eta}$
contains $\nu$.

Operation 5 checks in $O(\log |\ggg|)$ time whether a given type-4
$U_{i,j}$ for $\PT(\mu)$ is side-1, side-2, or two-sided for the minor
block $B_{k,l}$ in $\PT(\mu)$ strictly containing $U_{i,j}$.  Let
$\eta=\node(B_{k,l})$ and $\nu=\settle(U_{i,j})$.  Note that $\eta$
has been processed. If $\eta=\nu$, this operation takes $O(1)$ time
using the information stored for $\eta$. If $\nu$ is a descendent of
$\eta$, the representative $e$ of $\nu$ in $\SK(\eta)$ can be found in
$O(\log|\ggg|)$ time. Then, it takes $O(\log|\ggg|)$ time to check
whether $e$ is on $\side_1(\SK(\eta))$ or $\side_2(\SK(\eta))$ using
the information stored for $\eta$.

\begin{lemma}\label{fac:final}\

\begin{enumerate}
\item\label{famPart} $\{\EF(\mu)~|~\mu$ is a P-node or R-node\} is a
	partition of $\{\CC_1, \ldots, \CC_q\}$.
\item\label{subPart} $\{\ES(\mu)~|~\mu$ is a P-node or R-node\} is
	a partition of $\CC_1\cup \cdots\cup \CC_q$.
\item\label{famTimes} Each input family $\CC_i$ is processed exactly once.
\item\label{subTimes} Each input $U_{i,j}$ is processed at most twice,
   and the total time spent on processing $U_{i,j}$ is
   $O(|U_{i,j}|+\log|\ggg|)$.
\end{enumerate}
\end{lemma}
\begin{proof}
Statements \ref{famPart} and \ref{subPart} are straightforward.
Statement~\ref{famTimes} holds since each $\CC_i$ is processed only
when the node $\mu$ with $\CC_i\in\EF(\mu)$ is processed.  Each
$U_{i,j}$ is processed once when the node $\mu$ with $U_{i,j} \in
\ES(\mu)$ is processed and once when the node $\phi$ with $\CC_i \in
\EF(\phi)$ is processed. When $U_{i,j}$ is processed, we perform some
of Operations 1 through 5 on it. Since an operation takes
$O(|U_{i,j}|+\log|\ggg|)$ time, Statement~\ref{subTimes} holds.
\end{proof}

We now bound the time of processing an R-node or P-node $\mu$. Let
$\ESK(\mu)$ be obtained from $\SK(\mu)$ by replacing the virtual edge
of each S-child $\chi_k$ of $\mu$ with $\SK(\chi_k)$. Let $n_{\mu}$ be
the number of vertices in $\ESK(\mu)$. Let $N_{\mu}=\sum_{\CC_i \in
\EF(\mu)} |\CC_i|$.  Recall that $\mu$ is processed using some of the
following operations:
\begin{enumerate}
\item Process the sets $U_{i,j}$ in the families $\CC_i\in\EF(\mu)$.
\item Call Theorem~\ref{thm_m3}(\ref{thm_m3_2}) on input $\SK(\mu)$
      and $\MM'(\mu)$.
\item Call the S-procedure on $\chi_k$ for the S-children $\chi_k$ of
      $\mu$.
\item Compute $\ext_p(\PT(\mu))$ for $p=0$, 1, and 2.
\item 
Construct auxiliary graphs $D$, $RD$ and $\hh$, and operate on them.
\end{enumerate}
Note that each $K \in \{D,RD,\hh\}$ is constructed and operated on
in $O(|K|)$ total time.  Since $\sum_K |K| \leq n_{\mu}$ where $K$
ranges over all auxiliary graphs constructed during the processing of
$\mu$, it takes $O(n_{\mu})$ total time to process the auxiliary
graphs for $\mu$.  Therefore, the above operations take
$O((n_{\mu}+N_{\mu})\log\inputsize)$ time in total.  By summing over all
P-nodes and R-nodes $\mu$ of $T$, and by Theorem~\ref{thm_m3},
Fact~\ref{lemma:linear-space}(\ref{lemma:linear-space-1}), and
Lemma~\ref{fac:final}, the CFE algorithm runs in the desired total
time, completing the proof of Theorem \ref{thm_m2}.

\section{Directions for further research}\label{sec_open}
We have proved that the CFE problem can be solved in $O(\inputsize\log
\inputsize)$ time for the special case where for each input family
${\CC}_i$, each set in ${\CC}_i$ induces a connected subgraph of the
input graph $\ggg$. One direction for further research would be to
reduce the running time to linear. Such a result might lead to 
substantial simplification of the SPQR decomposition or an entirely
different data structure.  Another worthy direction would be to solve
more general cases in similar time bounds.  Beyond these technical
open problems, it would be of significance to find further
applications of the CFE problem than VLSI layout and topological
inference as well as to identify novel and fundamental constrained
planar embeddings.

\section*{Acknowledgments}
We wish to thank the anonymous referee for many helpful suggestions.

\bibliographystyle{abbrv}
\bibliography{all}
%\bibliography{kao}

\end{document}